\documentclass[footinbib,a4paper,aps,superscriptaddress,reprint,twocolumn,preprintnumbers,amsmath,amssymb,nobalancelastpage,10pt]{revtex4-2}

\usepackage{amsmath}
\usepackage{verbatim}
\usepackage{graphicx}
\usepackage{wasysym}

\usepackage{color}
\usepackage[dvipsnames]{xcolor}
\usepackage{braket}
\usepackage{yfonts}
\usepackage{soul}
\usepackage[normalem]{ulem}
\usepackage{dsfont}
\usepackage[colorlinks=true,citecolor=blue,linkcolor=blue,urlcolor=blue]{hyperref}

\usepackage{amsmath}
\usepackage{bm}
\usepackage{amsfonts}

\usepackage{graphicx}
\graphicspath{{Figures_paper}}

\begin{document}

\def \IBK{Institute for Theoretical Physics, University of Innsbruck, 6020 Innsbruck, Austria}
\def \IQOQI{Institute for Quantum Optics and Quantum Information of the Austrian Academy of Sciences, 6020 Innsbruck, Austria}

\title{Deconfined Quantum Criticality on a Triangular Rydberg Array}

\author{Lisa Bombieri}\email{lisa.bombieri@uibk.ac.at}\affiliation{\IBK}\affiliation{\IQOQI}
\author{Torsten V. Zache}\affiliation{\IBK}\affiliation{\IQOQI}
\author{Gabriele Calliari}\affiliation{\IBK}\affiliation{\IQOQI}
\author{Mikhail D. Lukin}
\affiliation{Department of Physics, Harvard University, Cambridge, MA 02138, USA}
\author{Hannes Pichler }\affiliation{\IBK}\affiliation{\IQOQI}
\author{Daniel Gonz\'alez-Cuadra}\email{daniel.gonzalez@ift.csic.es}
\affiliation{\IBK}
\affiliation{\IQOQI}
\affiliation{Department of Physics, Harvard University, Cambridge, Massachusetts 02138, USA}
\affiliation{Instituto de F\'isica Te\'orica UAM-CSIC, Calle Nicol\'as Cabrera 13-15, Cantoblanco, 28049 Madrid, Spain}

\begin{abstract}
Fluctuations can drive continuous phase transitions between two distinct ordered phases---so-called deconfined quantum critical points (DQCPs)---which lie beyond the Landau-Ginzburg-Wilson paradigm. Despite several theoretical predictions over the past decades, experimental evidence of DQCPs remains elusive. We show that a DQCP can be explored in a system of Rydberg atoms arranged on a triangular lattice and coupled through van der Waals interactions. Specifically, we investigate the nature of the phase transition between two ordered phases at $1/3$ and $2/3$ Rydberg excitation density, which were recently probed experimentally in [P. Scholl et al., Nature 595, 233 (2021)]. Using a field-theoretical analysis, we predict both the critical exponents for infinitely long cylinders of increasing circumference and the emergence of a conformal field theory near criticality showing an enlarged U($1$) symmetry---a signature of DQCPs---and confirm these predictions numerically.  
Finally, we extend these results to ladder geometries and show how the emergent U($1$) symmetry could be probed experimentally using finite tweezer arrays. 
\end{abstract}

\maketitle

\paragraph*{Introduction---}

Rydberg atoms trapped in optical tweezers~\cite{Browaeys_2020, Kaufman_2021} have emerged in recent years as a powerful platform for quantum simulation~\cite{Altman_2021}, enabling the exploration of quantum many-body physics in a highly controllable setting~\cite{Labuhn_2016, Bernien_2017, Leseleuc_2019, Keesling_2019, Ebadi_2021, Scholl_2021, Bluvstein_2021, Semeghini_2021, Choi_2023, Chen_2024, Shaw_2024, Manovitz_2024, Gonzalez-Cuadra_2024, Fang_2024}. In particular, the ability to arrange atoms in arbitrary geometries, combined with long-range Rydberg interactions, makes these systems especially well suited for investigating frustrated spin models~\cite{Sachdev_2023}. These models are characterized by many competing classical configurations, a setting where quantum fluctuations can give rise to long-range entangled states~\cite{Wen_2007}, as exemplified by the recent experiments probing a topological  spin-liquid phase~\cite{Semeghini_2021}.

The interplay between competing orders and quantum fluctuations can also lead to deconfined quantum critical points (DQCPs)~\cite{Senthil_2004, Senthil_2004_PRB, Shao_2016}, which correspond to continuous phase transitions between two distinct ordered phases, and that lie beyond the conventional Landau-Ginzburg-Wilson (LGW) paradigm~\cite{Landau_1980, Wilson_1974}.  DQCPs are characterized by emergent fractionalized excitations and deconfined gauge fields and have been predicted in a variety of models~\cite{Senthil_2024}. Despite significant theoretical and numerical efforts over the past decades, experimental evidence for DQCPs remains scarce~\cite{Zayed_2017, Guo_2020, Hong_2022, Cui_2023}. Identifying experimental platforms capable of probing this phenomenon directly is, therefore, of great interest.

In this work, we show that deconfined quantum criticality can be explored  in a system of Rydberg atoms arranged on a triangular lattice and coupled via van der Waals interactions. Such systems have already been realized experimentally~\cite{Scholl_2021}, where the ordered phases at $1/3$ and at $2/3$ filling were adiabatically prepared. In the region between these two phases, however, it is unclear whether the system hosts an intermediate phase induced by the order-by-disorder mechanism~\cite{Moessner_2000, Humeniuk_2016, Saadatmand_2018, Koziol_2019, Fey_2019} or instead features a first-order or continuous phase transition. For a point in parameter space lying in between the two ordered phases, Ref.~\cite{Guo_2023} found a finite-temperature Kosterlitz-Thouless (KT) phase---characterized by an emergent U($1$) symmetry. Here, we connect the emergent U($1$) symmetry to a DQCP---a continuous phase transition---separating the two distinct ordered phases observed in Ref.~\cite{Scholl_2021}.

To this end, we first study the model on infinitely long cylinders of increasing circumference. Combining field-theoretical analysis with large-scale density-matrix renormalization group (DMRG) simulations~\cite{Schollwock_2011}, we demonstrate the presence of a DQCP in the ground-state phase diagram of these quasi-1D systems. As in previous studies of DQCPs in 1D~\cite{Jiang_2019, Jiang_2019_2, Mudry_2019, Huang_2019, Zhang_2023, Lee_2023, Romen_2024, Baldelli_2024}, the critical point is described by a conformal field theory (CFT) with central charge $c = 1$ and an emergent U($1$) symmetry. We interpret the cylinder circumference as an inverse temperature of the corresponding 2D model and, using a dimensional reduction argument, show that our results are consistent with a DQCP over a finite temperature range---connecting our results with those found in Ref.~\cite{Guo_2023}. Furthermore, we extend our analysis to experimentally accessible ladder geometries, where we also find a DQCP in the ground state. Finally, we describe how the emergent U($1$) symmetry could be directly probed in Rydberg experiments via measurements in the occupation basis.

\begin{figure}[t]
    \centering
    \includegraphics[width=\columnwidth]{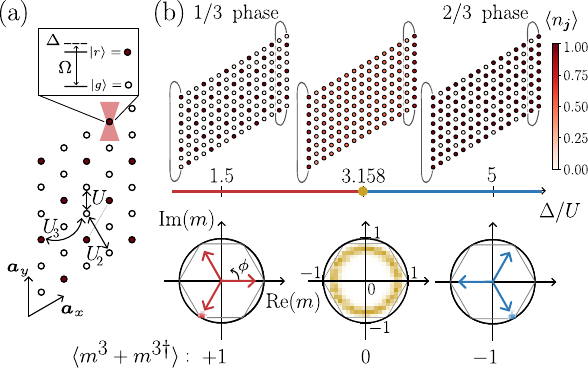}
    \caption{(a) Neutral atoms trapped on a triangular tweezer array with  lattice vectors ${\bf a}_x$ and ${\bf a}_y$, where we depict the laser-coupled electronic levels. (b) Ground-state phase diagram for a cylinder with $N_x=15$ atoms along the $x$ axis and $N_y=9$ atoms along the periodic $y$ axis. (Top) Average occupation number $\langle n_{\bf j} \rangle$ in the $1/3$ phase ($\Delta/U=1.5$), at the DQCP [$(\Delta/U)_{\rm c}=3.158$], and in the $2/3$ phase ($\Delta/U=5$). (Bottom) The $1/3$ ($2/3$) phase spontaneously breaks the $\mathbb{Z}_3$ symmetry: $m=|m|e^{i\phi}$ can point along $\phi=0,2\pi/3,4\pi/3$ ($\phi=\pi/3, \pi, 5\pi/3$), corresponding to a real order parameter $\langle m^3+m^{3\dagger}\rangle=+1(-1)$. At the transition, the $\mathbb{Z}_6$ symmetry is enlarged to U($1$). The histograms show the distribution of  $m$ [Eq.~\eqref{eq:magnetization}] over $10^4$ snapshots in the occupation basis when restricting the sites to the ``bulk", i.e.,  excluding the first and the last three rows of atoms along ${\bf a}_x$. 
    Results obtained using a finite MPS with bond dimension $\chi=100$ for $\Delta/U=1.5$ and $\Delta/U=5$, and $\chi=500$ for $\Delta/U=3.158$.}
    \label{fig:Fig1}
\end{figure}

\paragraph*{Model and results---} We consider neutral atoms trapped in optical tweezers arranged on a triangular lattice with basis vectors ${\bf a}_x = (\sqrt{3}/2,1/2)a $ and ${\bf a}_y=(0,1)a$, where $a$ is the lattice constant [Fig.~\ref{fig:Fig1}(a)]. 
The atoms are driven by a coherent laser field with Rabi frequency $\Omega$ and detuning $\Delta$, which couples two internal electronic states---the ground state $\ket{g}$ and a highly excited Rydberg level $\ket{r}$. This system is described by the following Hamiltonian~\cite{Scholl_2021}:
\begin{equation}
    H_{\rm Ryd}= \frac{\Omega}{2} \sum_{{\bf j}} \sigma^{x}_{{\bf j}} -\Delta \sum_{{\bf j}} n_{{\bf j}} + \sum_{{\bf i} > {\bf j}} U_{{\bf i} {\bf j}}\, n_{{\bf i}} n_{{\bf j}},
    \label{eq:hamiltonain_rydberg}
\end{equation}
where $n_{{\bf j}}=\ket{r}_{{\bf j}}\bra{r}$, $\sigma^{x}_{{\bf j}}=\ket{r}_{{\bf j}}\bra{g}+\ket{g}_{{\bf j}}\bra{r}$, and $U_{{\bf i} {\bf j}}=C_6/|x_{\bf i}-x_{\bf j}|^6$ is the van der Waals interaction strength between excited atoms at positions $x_{\bf i}$ and $x_{\bf j}$. In the following, we indicate the number of atoms along ${\bf a}_x$ (${\bf a}_y$) with $N_x$ ($N_y$), and the interaction strength between nearest neighbors with $U$.

In this work, we study the phase transition between the ordered phases at $1/3$ and $2/3$ filling~\cite{Scholl_2021}---hereafter referred to as the $1/3$ and $2/3$ phases for brevity---characterized by a nonzero staggered magnetization, 
\begin{equation}
    m = \frac{1}{N/3} \sum_{{\bf j}} n_{{\bf j}}\, e^{i {\bf Q} \cdot   x_{{\bf j}}},
    \label{eq:magnetization}
\end{equation}
with ${\bf Q}=\left(2\sqrt3/3,2/3\right)\pi/a$ and $x_{\bf j}$ being the position of the $\bf{j}$ atom. In the $1/3$ phase, the complex order parameter $m=|m|e^{i\phi}$ can point along $\phi=0,2\pi/3,$ or $4\pi/3$, while in the $2/3$ phase, it can point along $\phi=\pi/3, \pi,$ or $ 5\pi/3$ [Fig.~\ref{fig:Fig1}(b)].  The two ordered phases thus correspond to a spontaneous symmetry breaking of two distinct $\mathbb{Z}_3$ symmetries. In our work, to distinguish between the two, we use the real-valued order parameter $\langle m^3+m^{3 \dagger} \rangle = |m|^3 \cos{(3\phi)}$, which takes positive and negative values in the $1/3$ and $2/3$ phases. Since these phases are related by a $\mathbb{Z}_2$ inversion symmetry ($\ket{g}\to\ket{r}$),  a transition is expected at the point of degeneracy $(\Delta/U)_{\rm c}$, where the system exhibits a $\mathbb{Z}_3\times \mathbb{Z}_2\simeq \mathbb{Z}_6$ symmetry. 

In the recent study of Ref.~\cite{Guo_2023}, the authors focus on the critical point $(\Delta/U)_{\rm c}$ at fixed $\Omega/U\approx0.33$, analyzing the phase transition as a function of temperature. They show that thermal fluctuations render the $\mathbb{Z}_6$ anisotropy irrelevant, resulting in an extended KT phase with quasi-long-range order and emergent continuous U($1$) symmetry, separated from the high-temperature disordered phase by a KT transition.

\begin{figure*}
    \centering
    \includegraphics[width=\textwidth]{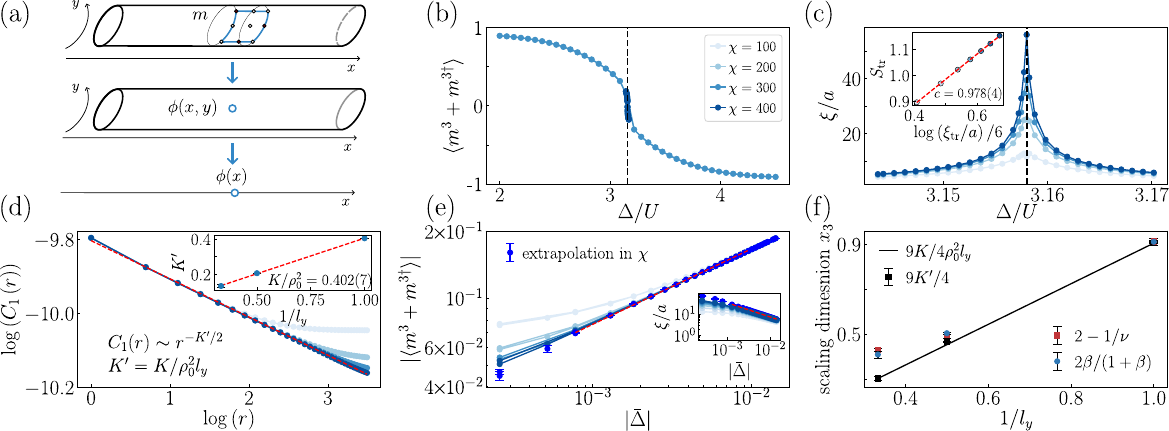}
    \caption{(a) Mapping from a 2D cylindrical lattice to a 1D field theory model. The phase of the magnetization $m$ [Eq.~\eqref{eq:magnetization}] in each unit cell (of size $3\times 3$) is mapped to a 2D field $\phi(x,y)$ and by dimensional reduction to a 1D field $\phi(x)$. (b)--(e) Analysis of an infinitely long cylinder ($N_x=\infty$) with circumference $N_y=6$  ($l_y=2$ unit cells) for increasing bond dimension $\chi$ [see legend in panel (b)]. (b) The order parameter changes continuously from positive to negative as a function of $\Delta/U$, passing through zero at the critical point at $(\Delta/U)_{\rm c}\approx 3.158$ (dashed vertical line). (c) The correlation length $\xi/a$ as a function of $\Delta/U$ shows a peak at $(\Delta/U)_{\rm c}$, which diverges with $\chi$. Inset: extraction of the central charge $c$ from the scaling $S_{\rm tr}=c\log{(\xi_{\rm tr}/a)}/6$ at $(\Delta/U)_{\rm c}$. (d) Power-law decay of the two-point correlation function $\langle m(r) m(0)\rangle \approx C_1(r)=\rho^2_0 \langle e^{i\phi(r)}  e^{-i\phi(0)} \rangle$ at $(\Delta/U)_{\rm c}$, where $m$ is evaluated on unit cells of size $3\times 3\,l_y$. Inset: Luttinger parameter $K'$ as a function of the cylinder circumferences $l_y$, extracted from the power-law decay of $C_1(r)$ [Table~\ref{tab:scaling_exponents}]. (e) Linear fit of the order parameter and the correlation length (inset) in log-log scale as a function of $\bar{\Delta}=\Delta/U-(\Delta/U)_{\rm c}$, used to extract the critical exponents $\beta$ and $\nu$, respectively. (f) Dependence of the critical exponents on $l_y$.  We compare the scaling dimension $x_3$ estimated from (red) $2-1/\nu$ and (blue) $2\beta/(1+\beta)$ to the theoretical prediction (black dots) $x_3= 9 K'/4$  and (black line) $x_3=9 K/(4 \rho^2_0l_y)$, with coefficient $K/\rho^2_0$ obtained from the fit in the inset of (d). }
    \label{fig:Fig2}
\end{figure*}

\begin{figure}[t]
    \centering
\includegraphics[width=\columnwidth]{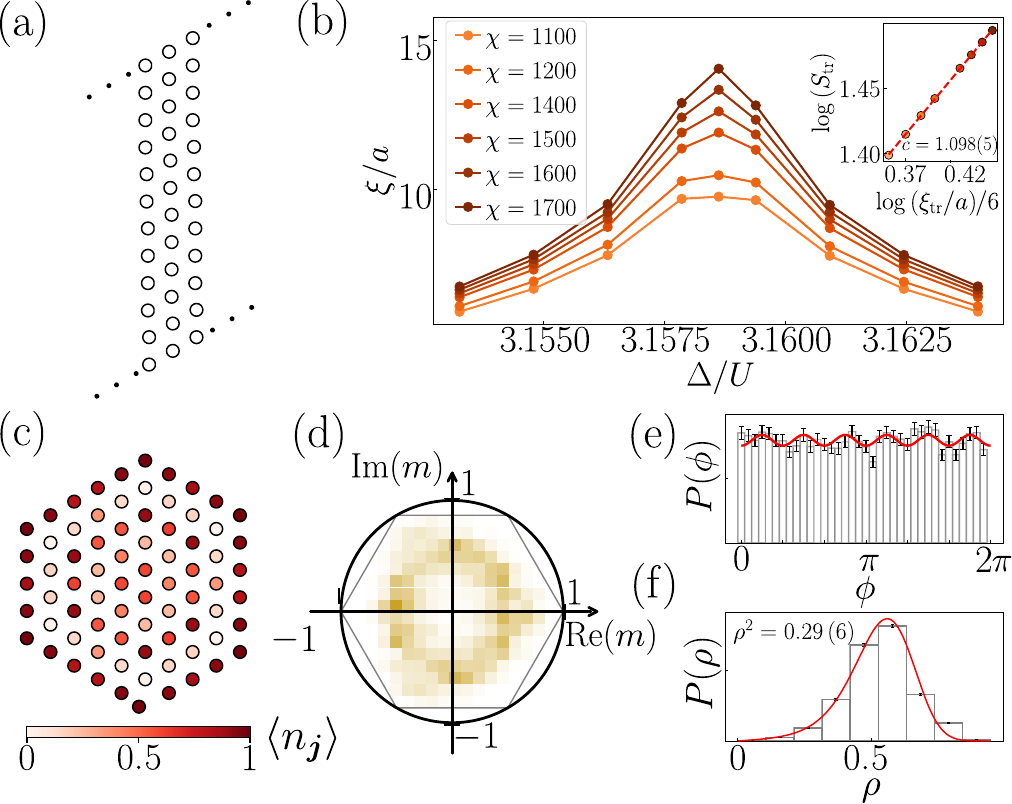}

    \caption{(a) Unit cell of the infinite ladder ($N_x=\infty$) with $N_y=3l_y=12$ atoms along ${\bf a}_y$. (b) Correlation length  $\xi/a$ as a function of $\Delta/U$ for the infinite ladder shown in (a) and increasing bond dimension $\chi$. Inset: extraction of the central charge $c$ from the scaling of $S_{\rm tr}=c\log{(\xi_{\rm tr}/a)}/6$ at $(\Delta/U)_{\rm c} \approx 3.1586$, corresponding to the maximum of the correlation length. (c)-(f) Experimentally feasible system. (c) Lattice and occupation number $\langle n_{\bf j} \rangle$ for $\chi=200$ and $\Delta/U=2.94$. (d) Distribution  of the staggered magnetization $m$ [Eq.~\eqref{eq:magnetization}] of the state considered in (c) over $10^4$ snapshots in the occupation basis when restricting the sites to the ``bulk" (i.e., excluding the last two ``rings" of atoms). (e) Angular distribution of $m$, consistent with a weak potential $V(\phi)$ due to finite size effects~\cite{SM}. (f) Radial distribution of $m$, consistent with a potential $V(\rho)$.}
    \label{fig:Fig3}
\end{figure}

In our work, we focus instead on the zero-temperature quantum phase transition between the $1/3$ and the $2/3$ phases as a function of $\Delta/U$ at fixed $\Omega/U=0.33$. 
Unlike Ref.~\cite{Guo_2023}, which considers an isotropic 2D system, we study quasi-1D geometries: infinite cylinders (Fig.~\ref{fig:Fig2}) and infinite ladders [Figs.~\ref{fig:Fig3}(a)-\ref{fig:Fig3}(b)]. Specifically, the cylinders have periodic boundary conditions along ${\bf a}_y$, with circumference $N_y$---corresponding to $l_y$ unit cells (of size $3\times 3$) along ${\bf a}_y$. These geometries allow us to characterize the transition theoretically via an effective field theory, and we expect them to be effectively equivalent to isotropic systems at finite temperatures---connecting with the results in Ref.~\cite{Guo_2023}--- set by the inverse of the cylinder circumference, i.e., $k_{\rm B}T\sim1/l_y$~\cite{SM}. Infinite ladders, by contrast, have open
boundary conditions along ${\bf a}_y$ and are infinite along ${\bf a}_x$, making them relevant to experimental Rydberg arrays. In our work, we show that in both cases the two ordered phases are separated by a DQCP with emergent U($1$) symmetry [see Fig.~\ref{fig:Fig1}(b)]. Finally, we also propose a finite-size array that can be experimentally realized, in which the emergent U($1$) symmetry can be observed [Figs.~\ref{fig:Fig3}(c)–\ref{fig:Fig3}(f)].

In the following, we truncate the interactions to the third-nearest neighbors [Fig.~\ref{fig:Fig1}(a)]. This approximation is justified for a small Rydberg radius---the characteristic distance where the interaction strength is dominant and satisfies $U_{{\bf i} {\bf j}} = \Omega$. For $\Omega/U=0.33$, this yields $R_{\rm b}=1.203a$, supporting the neglect of interactions beyond a distance of $2a$. In Supplementary Material (SM)~\cite{SM}, we show how our results, in particular the presence of a DQCP with an emergent U($1$) symmetry, remain unaffected by the presence of longer-range interaction terms and persist for a broad range of parameters up to $\Omega/U\approx0.4$.

\paragraph*{Effective field theory---}

We aim to characterize the transition on infinitely long cylinders with increasing circumference and to develop a field theory that both captures the nature of the transition and connects the results across different cylinders. 
We begin by presenting the results for the infinitely long cylinder with circumference $N_y=6$ ($l_y=2$). Further details on other geometries can be found in SM~\cite{SM}. Here, the order parameter changes continuously from positive ($1/3$ phase) to negative ($2/3$ phase), vanishing at the transition point $(\Delta/U)_{\rm c}\approx3.158$ [Fig.~\ref{fig:Fig2}(b)]. This smooth behavior signals a continuum quantum phase transition beyond the conventional LGW paradigm. This is further corroborated by the growth of the correlation length $\xi$ as the bond dimension $\chi$ increases in our DMRG simulations [Fig.~\ref{fig:Fig2}(c)]. In particular, at the critical point, the scaling of the von Neumann entanglement entropy at half cylinder with the correlation length, $S_{\rm tr}=c\log{(\xi_{\rm tr}/a)}/6$, is consistent with the emergence of a CFT with central charge $c=1$~\cite{Pasquale_Calabrese_2004, Tagliacozzo_2008, Pollmann_2009}. Further evidence for critical behavior is provided by the power-law decay of the two-point correlation function of the magnetization between unit cells separated by distance $r$, $\langle m(r) m(0)\rangle$ [Fig.~\ref{fig:Fig2}(d)]. 

To capture these numerical findings, we formulate an effective field theory by mapping the discrete microscopic model onto a continuous field description [Fig.~\ref{fig:Fig2}(a)]. In the limit of quasi-1D systems ($l_y\ll l_x$), we associate a complex field $\psi(x)=\rho(x)e^{i\phi(x)}$ to the staggered magnetization $m$ evaluated in a unit-cell of size $3\times N_y=3\times 3\,l_y$.  Neglecting  amplitude fluctuations, and setting $\psi(x)=\rho_0 e^{i\phi(x)}$, we can effectively describe the system via the 1D theory~\cite{SM}:
\begin{equation}
    S= \int dt \, dx \left[ \frac{\left(\partial_\mu \phi\right)^2}{2\pi K'}  + g'_3 \cos{\left(3 \phi\right)} +g'_6\cos{(6\phi)} \right],
\label{eq:effective_th_phase_0}
\end{equation}
where the parameters $K'$, $g'_3$, and $g'_6$ depend on $l_y$ and $\rho_0$. In particular, $K' =K/(\rho_0^2l_y)$, where $K$ is a constant.
This theory captures the phase diagram of the Rydberg model in Eq.~\eqref{eq:hamiltonain_rydberg} as a function of $\Delta/U$ (which we assume controls $g'_3$) and temperature $T$ (proportional to $1/l_y$ and thus to $K'$)~\cite{SM}. In particular, it allows for a quantitative characterization of the continuous transition between the two $\mathbb{Z}_3$-ordered phases, including predictions for the critical exponents, as detailed below.

Here, we focus on the regime $2/9< K'< 8/9$, where the $\mathbb{Z}_6$ anisotropy is irrelevant under the renormalization-group flow, while the $\mathbb{Z}_3$ anisotropy is relevant. A detailed discussion of the remaining parameter regimes is provided in SM~\cite{SM}. 
In this range, the system spontaneously breaks the $\mathbb{Z}_3$ symmetry: $\cos{(3\phi)}=1$ for $g'_3<0$, and $\cos{(3\phi)}=-1$ for $g'_3>0$, corresponding to the $1/3$ and $2/3$ phases, respectively. At the critical point, i.e., $g'_3=0$, the model reduces to a Luttinger liquid with an emergent U($1$) symmetry. 
Therefore, as $g'_3$ changes sign, the system undergoes an unconventional continuous phase transition between two distinct ordered phases, which are separated by a DQCP with U($1$) symmetry. We further predict the dependence of the critical exponents on the cylinder width $l_y$, as summarized in Table~\ref{tab:scaling_exponents}~\cite{SM}.
\begin{table}
    \centering
    \renewcommand{\arraystretch}{1.9}
    \begin{tabular}{|c|c|c|}
        \hline
        Observable & Scaling & Exponent  \\ \hline
        $C_n(r)$ & $r^{-2x_n}$ & $x_n =  \frac{n^2 K'}{4} = \frac{n^2 K}{4 \rho^2_0l_y}$ \\ \hline
        $\xi$ & ${|g_3'|}^{-\nu}$ & $\nu = \frac{1}{2 - x_3}$ \\ \hline
        $\rho^3_0|\langle \cos (3\phi) \rangle|$ & ${|g_3'|}^\beta$ & $\beta = \frac{x_3}{2 - x_3}$ \\ \hline
    \end{tabular}
    \caption{Correlation function and critical exponents for the field theory in Eq.~\eqref{eq:effective_th_phase_0} in the regime $2/9<K'<8/9$. The two-point correlation function $C_n(r) = \rho^2_0 \langle e^{in\phi(r)} e^{-in\phi(0)} \rangle$ at criticality decays algebraically with distance $r$. The correlation length $\xi$ and the order parameter $\rho^3_0 \langle \cos{(3\phi)} \rangle$ scale as power laws in the perturbation parameter $g'_3$. The associated critical exponents depend on the cylinder circumference $l_y$ through the effective Luttinger parameter $K'= K/(\rho_0^2l_y)$. }
    \label{tab:scaling_exponents}
\end{table}

We numerically verify the validity of this 1D model in Figs.~\ref{fig:Fig2}(d)-\ref{fig:Fig2}(f). First, we compute the two-point correlation function of the staggered magnetization evaluated for unit cells of size $3\times 3 \,l_y$ (with $l_y=1,2$, and $3$ in units of $3a$) as a function of distance $\langle m(r) m(0)\rangle$, which corresponds to the field-theoretical correlation $C_1(r) = \rho_0^2\langle e^{i\phi(r)} e^{-i\phi(0)}\rangle$. From its algebraic decay, we extract the effective Luttinger parameter $K'$ and find that, as predicted, it displays a linear dependence on $1/l_y$ [Fig.~\ref{fig:Fig2}(d)]. 

The critical exponents $\nu$  and $\beta$ are related via the scaling dimension $x_3=9K'/4$ of the perturbation $\cos{(3\phi)}$, according to the relation  $x_3=2-1/\nu = 2 \beta/(1+\beta)$.
We test this relation by extracting the critical exponents $\nu$ and $\beta$ from fits of the correlation length and order parameter, respectively, as a function of the distance from the critical point, $\bar{\Delta}=\Delta/U-(\Delta/U)_{\rm c}$, which plays the role of the perturbation strength $g'_3$, i.e.,  $ \bar{\Delta} \propto g'_3$ [Fig.~\ref{fig:Fig2}(e)]. The exponents obtained for different circumferences $l_y$ agree with the theoretical predictions based on the Luttinger parameter $K'$ for small $l_y$, with deviations that increase as $l_y$ becomes larger [Fig.~\ref{fig:Fig2}(f)]. In SM~\cite{SM}, we further corroborate the effective model on finite cylinders by performing a finite-size scaling analysis with respect to the cylinder length and analyzing the statistics of the staggered magnetization near criticality.

Finally, we note that the $\mathbb{Z}_6$ anisotropy becomes relevant in the 1D model when $K'<2/9$, a condition met for large $l_y$. In particular, for $l_y=3$, we numerically find $K'\sim 0.1$ [Fig.~\ref{fig:Fig2}(d)]. The deviations observed for this cylinder may therefore originate either from the growth of the $\mathbb{Z}_6$ anisotropy or from the breakdown of the effective 1D description.
The relevance of the $\mathbb{Z}_6$ anisotropy could be inferred from the saturation of the correlation length $\xi$ as the bond dimension $\chi$ increases at criticality. Indeed, when relevant, the anisotropy leads---depending on the sign of $g'_6$---either to a first-order transition or to an intermediate phase with spontaneously broken $\mathbb{Z}_6$ symmetry separating the $1/3$ and the $2/3$ phases~\cite{SM}. However, our numerical results for $l_y=3$ show that $\xi$ continues to grow with increasing $\chi$, albeit at a slower rate, suggesting that the system remains critical and that the 1D approximation may be breaking down. To definitely rule out the relevance of the $\mathbb{Z}_6$ anisotropy, simulations at larger bond dimensions ($\chi>2000$) would be necessary. The fate of the transition in the infinite-circumference limit is of particular interest for quantum simulations on experimental platforms, as we discuss now.

\paragraph*{Finite ladders and experimental proposal---}

Apart from probing the fate of an isotropic 2D system at zero temperature, we now argue that current Rydberg experiments could be readily used to observe a DQCP in an accessible quasi-1D geometry. This also represents an interesting direction, as various experimental proposals have been advanced for observing DQCPs in 1D systems~\cite{Lee_2023, Romen_2024, Baldelli_2024}, yet none have been realized so far.
Previously, we have shown that the transition on quasi-1D cylinders is unconventional. However, realizing cylindrical geometries in experiments is challenging. To address this, we now consider ``infinite ladders"---systems that have open boundary conditions along ${\bf a}_y$ and are infinite along ${\bf a}_x$ [Fig.~\ref{fig:Fig3}(a)]. Open boundaries play an important role, giving rise to an intermediate phase for small ladder widths ($l_y<4$). However, for $l_y=4$ we recover the continuous transition between the $1/3$ and $2/3$ phases. In particular, at criticality, both the entanglement entropy and the correlation length increase with bond dimension, consistent with a CFT with $c=1$ [Fig.~\ref{fig:Fig3}(b)].

As an experimentally feasible setup to observe this DQCP, we consider the finite array shown in Fig.~\ref{fig:Fig3}(c), where the six classical configurations at $1/3$ and $2/3$ filling are exactly degenerate in energy in the limit $\Omega = 0$. We note that similar geometries were realized in Ref.~\cite{Scholl_2021}. Remarkably, even for such a small system, we find that an approximately U($1$)-symmetric angular distribution emerges between the $1/3$ and $2/3$ phases [Fig.~\ref{fig:Fig3}(d)], signaling the presence of deconfined quantum criticality. 
In particular, the angular probability distribution $P(\phi)$ and the radial probability distribution $P(\rho)$ agree with the field-theory predictions [Figs.~\ref{fig:Fig3}(e)-\ref{fig:Fig3}(f)] and remain unaffected by the presence of longer-range interaction terms~\cite{SM}. The distribution $P(\phi)$ is consistent with a weak potential of the form $V(\phi)=g'_6 \cos{(6\phi)}$, such that $P(\phi)\propto e^{-V(\phi)}$, while the distribution $P(\rho)$ is compatible with a simple local effective potential $V_{\rm eff}(\rho) = \mu \rho^2 - \lambda \rho^4$, with $\rho_0^2 = \mu/2\lambda$, such that $P(\rho)\propto e^{-V_{\rm eff}(\rho)}$~\cite{SM}.

Scaling up this geometry in one direction would allow for an experimental probe of the critical properties of the quasi-1D ladder depicted in Fig.~\ref{fig:Fig3}(a). In particular, the DQCP can be identified through the progressive suppression of oscillations in the angular distribution $P(\phi)$ as the system size is increased in one direction. Whether or not the emergent U($1$) symmetry is spontaneously broken in the 2D thermodynamic limit remains an open and interesting question that cannot be definitely answered from our numerics. This question could also be addressed experimentally by scaling up this geometry in both directions.

\paragraph*{Conclusion and outlook---}
In this work, we have shown that quantum fluctuations render the phase transition between the $1/3$ and $2/3$ ordered phases of a triangular Rydberg array continuous in quasi-1D geometries, for both periodic and open boundary conditions. The transition point corresponds to a DQCP, which cannot be captured by the standard LGW formalism. 
Furthermore, our results demonstrated that the emergence of a continuous U($1$) symmetry at criticality is captured by a 1D conformal field theory and can be experimentally probed in accessible finite geometries. This opens several exciting directions for future research.

An important open question is whether the DQCP persists down to zero temperature in isotropic 2D systems. This question could also be addressed experimentally---as Rydberg atom arrays enable access to system sizes and geometries beyond current numerical capabilities---by adiabatically preparing the U(1)-symmetric ground state at the transition in systems of increasing size. However, adiabatically preparing a critical state requires a preparation time that grows (polynomially) with system size, imposing limits due to finite coherence times and experimental imperfections (see discussion in SM~\cite{SM}).  Possible future directions include performing numerical analyses of adiabatic state preparation to identify optimal laser sweep protocols to prepare the U(1)-symmetric angular distribution demonstrated in Figs.~\ref{fig:Fig3}(b)–\ref{fig:Fig3}(e), as well as dynamically probing the DQCP via the quantum Kibble-Zurek mechanism~\cite{Keesling_2019, Huang_2020, Shu_2025}.
Tensor network calculations were performed using the TeNPy Library~\cite{tenpy2024}.

\let\oldaddcontentsline\addcontentsline
\renewcommand{\addcontentsline}[3]{}
\paragraph*{Acknowledgments---} D.G.-C. thanks Daniel Barredo and Vincent Lienhard for helpful discussions on the experiments performed in Ref.~\cite{Scholl_2021}. We thank Sasha Geim for helpful discussions on a possible experimental implementation. This work is supported by the European Union’s Horizon Europe research and innovation program under Grant Agreement No.~101113690 (PASQuanS2.1), the ERC Starting grant QARA (Grant No.~101041435), and by the Austrian Science Fund (FWF) (Grant No. DOI 10.55776/COE1).
D.G.-C. acknowledges financial support through the Ramón y Cajal Program (RYC2023-044201-I), financed by MICIU/AEI/10.13039/501100011033 and by the FSE+.  The results in Fig.~\ref{fig:Fig3} have been obtained using the LEO HPC infrastructure of the University of Innsbruck.

\paragraph*{Data availability---}The data that support the findings of
this article are openly available~\cite{data}.

\bibliography{biblio_arxiv}

@article{Scholl_2021,
	author = {Scholl, Pascal and Schuler, Michael and Williams, Hannah J. and Eberharter, Alexander A. and Barredo, Daniel and Schymik, Kai-Niklas and Lienhard, Vincent and Henry, Louis-Paul and Lang, Thomas C. and Lahaye, Thierry and L{\"a}uchli, Andreas M. and Browaeys, Antoine},
	date = {2021/07/01},
	doi = {10.1038/s41586-021-03585-1},
	id = {Scholl2021},
	isbn = {1476-4687},
	journal = {Nature},
	number = {7866},
	pages = {233--238},
	title = {{Quantum simulation of 2D antiferromagnets with hundreds of Rydberg atoms}},
	url = {https://doi.org/10.1038/s41586-021-03585-1},
	volume = {595},
	year = {2021}}

@article{Guo_2023,
  title = {{Order by disorder and an emergent Kosterlitz-Thouless phase in a triangular Rydberg array}},
  author = {Guo, Sibo and Huang, Juntao and Hu, Jiangping and Li, Zi-Xiang},
  journal = {Phys. Rev. A},
  volume = {108},
  issue = {5},
  pages = {053314},
  numpages = {15},
  year = {2023},
  month = {Nov},
  publisher = {American Physical Society},
  doi = {10.1103/PhysRevA.108.053314},
  url = {https://link.aps.org/doi/10.1103/PhysRevA.108.053314}
}

@article{Saadatmand_2018,
  title = {{Phase diagram of the quantum Ising model with long-range interactions on an infinite-cylinder triangular lattice}},
  author = {Saadatmand, S. N. and Bartlett, S. D. and McCulloch, I. P.},
  journal = {Phys. Rev. B},
  volume = {97},
  issue = {15},
  pages = {155116},
  numpages = {14},
  year = {2018},
  month = {Apr},
  publisher = {American Physical Society},
  doi = {10.1103/PhysRevB.97.155116},
  url = {https://link.aps.org/doi/10.1103/PhysRevB.97.155116}
}

@article{Fey_2019,
  title = {{Quantum Criticality of Two-Dimensional Quantum Magnets with Long-Range Interactions}},
  author = {Fey, Sebastian and Kapfer, Sebastian C. and Schmidt, Kai Phillip},
  journal = {Phys. Rev. Lett.},
  volume = {122},
  issue = {1},
  pages = {017203},
  numpages = {6},
  year = {2019},
  month = {Jan},
  publisher = {American Physical Society},
  doi = {10.1103/PhysRevLett.122.017203},
  url = {https://link.aps.org/doi/10.1103/PhysRevLett.122.017203}
}

@article{Senthil_2004,
	author = {T. Senthil and Ashvin Vishwanath and Leon Balents and Subir Sachdev and Matthew P. A. Fisher},
	doi = {10.1126/science.1091806},
    journal = {Science},
    volume={303},
	number = {5663},
	pages = {1490-1494},
	title = {{Deconfined Quantum Critical Points}},
	year = {2004}
}

@article{Senthil_2004_PRB,
  title = {{Quantum criticality beyond the Landau-Ginzburg-Wilson paradigm}},
  author = {Senthil, T. and Balents, Leon and Sachdev, Subir and Vishwanath, Ashvin and Fisher, Matthew P. A.},
  journal = {Phys. Rev. B},
  volume = {70},
  issue = {14},
  pages = {144407},
  numpages = {33},
  year = {2004},
  month = {Oct},
  publisher = {American Physical Society},
  doi = {10.1103/PhysRevB.70.144407}}

@article{Shao_2016,
	author = {Hui Shao and Wenan Guo and Anders W. Sandvik},
	journal = {Science},
	number = {6282},
	pages = {213-216},
	title = {{Quantum criticality with two length scales}},
	volume = {352},
	year = {2016}}

@article{Pollmann_2009,
  title = {{Theory of Finite-Entanglement Scaling at One-Dimensional Quantum Critical Points}},
  author = {Pollmann, Frank and Mukerjee, Subroto and Turner, Ari M. and Moore, Joel E.},
  journal = {Phys. Rev. Lett.},
  volume = {102},
  issue = {25},
  pages = {255701},
  numpages = {4},
  year = {2009},
  month = {Jun},
  publisher = {American Physical Society},
  doi = {10.1103/PhysRevLett.102.255701},
  url = {https://link.aps.org/doi/10.1103/PhysRevLett.102.255701}
}

@article{Pasquale_Calabrese_2004,
    doi = {10.1088/1742-5468/2004/06/P06002},
    year = {2004},
    month = {jun},
    publisher = {},
    volume = {2004},
    number = {06},
    pages = {P06002},
    author = {Pasquale Calabrese and John Cardy},
    title = {{Entanglement entropy and quantum field theory}},
    journal = {Journal of Statistical Mechanics: Theory and Experiment}
}

@article{Tagliacozzo_2008,
  title = {{Scaling of entanglement support for matrix product states}},
  author = {Tagliacozzo, L. and de Oliveira, Thiago. R. and Iblisdir, S. and Latorre, J. I.},
  journal = {Phys. Rev. B},
  volume = {78},
  issue = {2},
  pages = {024410},
  numpages = {14},
  year = {2008},
  month = {Jul},
  publisher = {American Physical Society},
  doi = {10.1103/PhysRevB.78.024410},
  url = {https://link.aps.org/doi/10.1103/PhysRevB.78.024410}
}

@article{Zayed_2017,
	author = {Zayed, M. E. and R{\"u}egg, Ch. and Larrea J. , J. and L{\"a}uchli, A. M. and Panagopoulos, C. and Saxena, S. S. and Ellerby, M. and McMorrow, D. F. and Str{\"a}ssle, Th. and Klotz, S. and Hamel, G. and Sadykov, R. A. and Pomjakushin, V. and Boehm, M. and Jim{\'e}nez--Ruiz, M. and Schneidewind, A. and Pomjakushina, E. and Stingaciu, M. and Conder, K. and R{\o}nnow, H. M.},
	date = {2017/10/01},
	doi = {10.1038/nphys4190},
	id = {Zayed2017},
	isbn = {1745-2481},
	journal = {Nature Physics},
	number = {10},
	pages = {962--966},
	title = {{4-spin plaquette singlet state in the Shastry--Sutherland compound ${\mathrm{SrCu}}_{2}({\mathrm{BO}}_{3}{)}_{2}$}},
	url = {https://doi.org/10.1038/nphys4190},
	volume = {13},
	year = {2017}}

@article{Guo_2020,
  title = {{Quantum Phases of ${\mathrm{SrCu}}_{2}({\mathrm{BO}}_{3}{)}_{2}$ from High-Pressure Thermodynamics}},
  author = {Guo, Jing and Sun, Guangyu and Zhao, Bowen and Wang, Ling and Hong, Wenshan and Sidorov, Vladimir A. and Ma, Nvsen and Wu, Qi and Li, Shiliang and Meng, Zi Yang and Sandvik, Anders W. and Sun, Liling},
  journal = {Phys. Rev. Lett.},
  volume = {124},
  issue = {20},
  pages = {206602},
  numpages = {6},
  year = {2020},
  month = {May},
  publisher = {American Physical Society},
  doi = {10.1103/PhysRevLett.124.206602},
  url = {https://link.aps.org/doi/10.1103/PhysRevLett.124.206602}
}

@article{Hong_2022,
	author = {Hong, Tao and Ying, Tao and Huang, Qing and Dissanayake, Sachith E. and Qiu, Yiming and Turnbull, Mark M. and Podlesnyak, Andrey A. and Wu, Yan and Cao, Huibo and Liu, Yaohua and Umehara, Izuru and Gouchi, Jun and Uwatoko, Yoshiya and Matsuda, Masaaki and Tennant, David A. and Chern, Gia-Wei and Schmidt, Kai P. and Wessel, Stefan},
	date = {2022/06/02},
	doi = {10.1038/s41467-022-30769-8},
	id = {Hong2022},
	isbn = {2041-1723},
	journal = {Nature Communications},
	number = {1},
	pages = {3073},
	title = {{Evidence for pressure induced unconventional quantum criticality in the coupled spin ladder antiferromagnet ${\mathrm{C}}_{9}{\mathrm{H}}_{18}{\mathrm{N}}_{2}{\mathrm{Cu}}{\mathrm{Br}}_{4}$ }}, 
	url = {https://doi.org/10.1038/s41467-022-30769-8},
	volume = {13},
	year = {2022}}

@article{Cui_2023,
    author = {Yi Cui  and Lu Liu  and Huihang Lin  and Kai-Hsin Wu  and Wenshan Hong  and Xuefei Liu  and Cong Li  and Ze Hu  and Ning Xi  and Shiliang Li  and Rong Yu  and Anders W. Sandvik  and Weiqiang Yu },
    title = {{Proximate deconfined quantum critical point in ${\mathrm{SrCu}}_{2}({\mathrm{BO}}_{3}{)}_{2}$}},
    journal = {Science},
    volume = {380},
    number = {6650},
    pages = {1179-1184},
    year = {2023},
    doi = {10.1126/science.adc9487},
    URL = {https://www.science.org/doi/abs/10.1126/science.adc9487}
}

@article{Jiang_2019,
  title = {{Ising ferromagnet to valence bond solid transition in a one-dimensional spin chain: Analogies to deconfined quantum critical points}},
  author = {Jiang, Shenghan and Motrunich, Olexei},
  journal = {Phys. Rev. B},
  volume = {99},
  issue = {7},
  pages = {075103},
  numpages = {31},
  year = {2019},
  month = {Feb},
  publisher = {American Physical Society},
  doi = {10.1103/PhysRevB.99.075103},
  url = {https://link.aps.org/doi/10.1103/PhysRevB.99.075103}
}

@article{Zhang_2023,
  title = {{Exactly Solvable Model for a Deconfined Quantum Critical Point in 1D}},
  author = {Zhang, Carolyn and Levin, Michael},
  journal = {Phys. Rev. Lett.},
  volume = {130},
  issue = {2},
  pages = {026801},
  numpages = {6},
  year = {2023},
  month = {Jan},
  publisher = {American Physical Society},
  doi = {10.1103/PhysRevLett.130.026801},
  url = {https://link.aps.org/doi/10.1103/PhysRevLett.130.026801}
}

@Article{Romen_2024,
	title={{Deconfined quantum criticality in the long-range, anisotropic Heisenberg chain}},
	author={Anton Romen and Stefan Birnkammer and Michael Knap},
	journal={SciPost Phys. Core},
	volume={7},
	pages={008},
	year={2024},
	publisher={SciPost},
	doi={10.21468/SciPostPhysCore.7.1.008},
	url={https://scipost.org/10.21468/SciPostPhysCore.7.1.008},
}

@article{Lee_2023,
  title = {{Landau-Forbidden Quantum Criticality in Rydberg Quantum Simulators}},
  author = {Lee, Jong Yeon and Ramette, Joshua and Metlitski, Max A. and Vuleti\ifmmode \acute{c}\else \'{c}\fi{}, Vladan and Ho, Wen Wei and Choi, Soonwon},
  journal = {Phys. Rev. Lett.},
  volume = {131},
  issue = {8},
  pages = {083601},
  numpages = {7},
  year = {2023},
  month = {Aug},
  publisher = {American Physical Society},
  doi = {10.1103/PhysRevLett.131.083601},
  url = {https://link.aps.org/doi/10.1103/PhysRevLett.131.083601}
}

@article{Baldelli_2024,
  title = {{Frustrated Extended Bose-Hubbard Model and Deconfined Quantum Critical Points with Optical Lattices at the Antimagic Wavelength}},
  author = {Baldelli, Niccol\`o and Cabrera, Cesar R. and Juli\`a-Farr\'e, Sergi and Aidelsburger, Monika and Barbiero, Luca},
  journal = {Phys. Rev. Lett.},
  volume = {132},
  issue = {15},
  pages = {153401},
  numpages = {8},
  year = {2024},
  month = {Apr},
  publisher = {American Physical Society},
  doi = {10.1103/PhysRevLett.132.153401},
  url = {https://link.aps.org/doi/10.1103/PhysRevLett.132.153401}
}

@article{Jiang_2019_2,
  title = {{Deconfined quantum critical point in one dimension}},
  author = {Roberts, Brenden and Jiang, Shenghan and Motrunich, Olexei I.},
  journal = {Phys. Rev. B},
  volume = {99},
  issue = {16},
  pages = {165143},
  numpages = {19},
  year = {2019},
  month = {Apr},
  publisher = {American Physical Society},
  doi = {10.1103/PhysRevB.99.165143},
  url = {https://link.aps.org/doi/10.1103/PhysRevB.99.165143}
}

@article{Huang_2019,
  title = {{Emergent symmetry and conserved current at a one-dimensional incarnation of deconfined quantum critical point}},
  author = {Huang, Rui-Zhen and Lu, Da-Chuan and You, Yi-Zhuang and Meng, Zi Yang and Xiang, Tao},
  journal = {Phys. Rev. B},
  volume = {100},
  issue = {12},
  pages = {125137},
  numpages = {16},
  year = {2019},
  month = {Sep},
  publisher = {American Physical Society},
  doi = {10.1103/PhysRevB.100.125137},
  url = {https://link.aps.org/doi/10.1103/PhysRevB.100.125137}
}

@article{Mudry_2019,
  title = {{Quantum phase transitions beyond Landau-Ginzburg theory in one-dimensional space revisited}},
  author = {Mudry, Christopher and Furusaki, Akira and Morimoto, Takahiro and Hikihara, Toshiya},
  journal = {Phys. Rev. B},
  volume = {99},
  issue = {20},
  pages = {205153},
  numpages = {22},
  year = {2019},
  month = {May},
  publisher = {American Physical Society},
  doi = {10.1103/PhysRevB.99.205153},
  url = {https://link.aps.org/doi/10.1103/PhysRevB.99.205153}
}

@inbook{Senthil_2024,
    author = {T. Senthil},
    title = {{Deconfined Quantum Critical Points: A Review}},
    booktitle = {50 Years of the Renormalization Group},
    chapter = {Chapter 14},
    pages = {169-195},
    year = {2024},
    publisher={World Scientific},
    doi = {10.1142/9789811282386_0014},
    URL = {https://www.worldscientific.com/doi/abs/10.1142/9789811282386_0014}
}

@book{Landau_1980,
  title        = {{Statistical Physics}},
  author       = {Landau, L. D. and Lifshitz, E. M.},
  year         = {1980},
  edition      = {3rd},
  publisher    = {Butterworth-Heinemann},
  series       = {Course of Theoretical Physics},
  volume       = {5},
  address      = {Oxford},
  isbn         = {978-0-08-057046-4},
  doi          = {10.1016/C2009-0-24487-4},
  url          = {https://www.sciencedirect.com/book/9780080570464/statistical-physics}
}

@article{Wilson_1974,
title = {{The renormalization group and the $\epsilon$ expansion}},
journal = {Physics Reports},
volume = {12},
number = {2},
pages = {75-199},
year = {1974},
issn = {0370-1573},
doi = {https://doi.org/10.1016/0370-1573(74)90023-4},
url = {https://www.sciencedirect.com/science/article/pii/0370157374900234},
author = {Kenneth G. Wilson and J. Kogut}
}

@article{Chen_2024,
	author = {Chen, Cheng and Bornet, Guillaume and Bintz, Marcus and Emperauger, Gabriel and Leclerc, Lucas and Liu, Vincent S. and Scholl, Pascal and Barredo, Daniel and Hauschild, Johannes and Chatterjee, Shubhayu and Schuler, Michael and L{\"a}uchli, Andreas M. and Zaletel, Michael P. and Lahaye, Thierry and Yao, Norman Y. and Browaeys, Antoine},
	date = {2023/04/01},
	doi = {10.1038/s41586-023-05859-2},
	id = {Chen2023},
	isbn = {1476-4687},
	journal = {Nature},
	number = {7958},
	pages = {691--695},
	title = {{Continuous symmetry breaking in a two-dimensional Rydberg array}},
	url = {https://doi.org/10.1038/s41586-023-05859-2},
	volume = {616},
	year = {2023}}

@article{Manovitz_2024,
	author = {Manovitz, Tom and Li, Sophie H. and Ebadi, Sepehr and Samajdar, Rhine and Geim, Alexandra A. and Evered, Simon J. and Bluvstein, Dolev and Zhou, Hengyun and Koyluoglu, Nazli Ugur and Feldmeier, Johannes and Dolgirev, Pavel E. and Maskara, Nishad and Kalinowski, Marcin and Sachdev, Subir and Huse, David A. and Greiner, Markus and Vuleti{\'c}, Vladan and Lukin, Mikhail D.},
	date = {2025/02/01},
	doi = {10.1038/s41586-024-08353-5},
	id = {Manovitz2025},
	isbn = {1476-4687},
	journal = {Nature},
	number = {8049},
	pages = {86--92},
	title = {{Quantum coarsening and collective dynamics on a programmable simulator}},
	url = {https://doi.org/10.1038/s41586-024-08353-5},
	volume = {638},
	year = {2025}}

@article{Semeghini_2021,
	author = {G. Semeghini and H. Levine and A. Keesling and S. Ebadi and T. T. Wang and D. Bluvstein and R. Verresen and H. Pichler and M. Kalinowski and R. Samajdar and A. Omran and S. Sachdev and A. Vishwanath and M. Greiner and V. Vuleti{\'c} and M. D. Lukin},
	doi = {10.1126/science.abi8794},
	journal = {Science},
	number = {6572},
	pages = {1242-1247},
	title = {{Probing topological spin liquids on a programmable quantum simulator}},
	url = {https://www.science.org/doi/abs/10.1126/science.abi8794},
	volume = {374},
	year = {2021}}

@article{Bernien_2017,
	author = {Bernien, Hannes and Schwartz, Sylvain and Keesling, Alexander and Levine, Harry and Omran, Ahmed and Pichler, Hannes and Choi, Soonwon and Zibrov, Alexander S. and Endres, Manuel and Greiner, Markus and Vuleti{\'c}, Vladan and Lukin, Mikhail D.},
	date = {2017/11/01},
	doi = {10.1038/nature24622},
	id = {Bernien2017},
	isbn = {1476-4687},
	journal = {Nature},
	number = {7682},
	pages = {579--584},
	title = {{Probing many-body dynamics on a 51-atom quantum simulator}},
	url = {https://doi.org/10.1038/nature24622},
	volume = {551},
	year = {2017}}

@article{Keesling_2019,
	author = {Keesling, Alexander and Omran, Ahmed and Levine, Harry and Bernien, Hannes and Pichler, Hannes and Choi, Soonwon and Samajdar, Rhine and Schwartz, Sylvain and Silvi, Pietro and Sachdev, Subir and Zoller, Peter and Endres, Manuel and Greiner, Markus and Vuleti{\'c}, Vladan and Lukin, Mikhail D.},
	date = {2019/04/01},
	doi = {10.1038/s41586-019-1070-1},
	id = {Keesling2019},
	isbn = {1476-4687},
	journal = {Nature},
	number = {7751},
	pages = {207--211},
	title = {{Quantum Kibble--Zurek mechanism and critical dynamics on a programmable Rydberg simulator}},
	url = {https://doi.org/10.1038/s41586-019-1070-1},
	volume = {568},
	year = {2019}}

@article{Bluvstein_2021,
	author = {D. Bluvstein and A. Omran and H. Levine and A. Keesling and G. Semeghini and S. Ebadi and T. T. Wang and A. A. Michailidis and N. Maskara and W. W. Ho and S. Choi and M. Serbyn and M. Greiner and V. Vuleti{\'c} and M. D. Lukin},
	doi = {10.1126/science.abg2530},
	journal = {Science},
	number = {6536},
	pages = {1355-1359},
	title = {{Controlling quantum many-body dynamics in driven Rydberg atom arrays}},
	url = {https://www.science.org/doi/abs/10.1126/science.abg2530},
	volume = {371},
	year = {2021}}

@article{Ebadi_2021,
	author = {Ebadi, Sepehr and Wang, Tout T. and Levine, Harry and Keesling, Alexander and Semeghini, Giulia and Omran, Ahmed and Bluvstein, Dolev and Samajdar, Rhine and Pichler, Hannes and Ho, Wen Wei and Choi, Soonwon and Sachdev, Subir and Greiner, Markus and Vuleti{\'c}, Vladan and Lukin, Mikhail D.},
	date = {2021/07/01},
	doi = {10.1038/s41586-021-03582-4},
	id = {Ebadi2021},
	isbn = {1476-4687},
	journal = {Nature},
	number = {7866},
	pages = {227--232},
	title = {{Quantum phases of matter on a 256-atom programmable quantum simulator}},
	url = {https://doi.org/10.1038/s41586-021-03582-4},
	volume = {595},
	year = {2021}}

@article{Gonzalez-Cuadra_2024,
	author = {Gonz{\'a}lez-Cuadra, Daniel and Hamdan, Majd and Zache, Torsten V. and Braverman, Boris and Kornja{\v c}a, Milan and Lukin, Alexander and Cant{\'u}, Sergio H. and Liu, Fangli and Wang, Sheng-Tao and Keesling, Alexander and Lukin, Mikhail D. and Zoller, Peter and Bylinskii, Alexei},
	date = {2025/06/01},
	doi = {10.1038/s41586-025-09051-6},
	id = {Gonz{\'a}lez-Cuadra2025},
	isbn = {1476-4687},
	journal = {Nature},
	number = {8067},
	pages = {321--326},
	title = {{Observation of string breaking on a (2 + 1)D Rydberg quantum simulator}},
	url = {https://doi.org/10.1038/s41586-025-09051-6},
	volume = {642},
	year = {2025}}

@misc{Fang_2024,
      title={Probing critical phenomena in open quantum systems using atom arrays}, 
      author={Fang Fang and Kenneth Wang and Vincent S. Liu and Yu Wang and Ryan Cimmino and Julia Wei and Marcus Bintz and Avery Parr and Jack Kemp and Kang-Kuen Ni and Norman Y. Yao},
      year={2024},
      eprint={2402.15376},
      archivePrefix={arXiv}
}

@article{Choi_2023,
	author = {Choi, Joonhee and Shaw, Adam L. and Madjarov, Ivaylo S. and Xie, Xin and Finkelstein, Ran and Covey, Jacob P. and Cotler, Jordan S. and Mark, Daniel K. and Huang, Hsin-Yuan and Kale, Anant and Pichler, Hannes and Brand{\~a}o, Fernando G. S. L. and Choi, Soonwon and Endres, Manuel},
	date = {2023/01/01},
	doi = {10.1038/s41586-022-05442-1},
	id = {Choi2023},
	isbn = {1476-4687},
	journal = {Nature},
	number = {7944},
	pages = {468--473},
	title = {Preparing random states and benchmarking with many-body quantum chaos},
	url = {https://doi.org/10.1038/s41586-022-05442-1},
	volume = {613},
	year = {2023}}

@article{Shaw_2024,
	author = {Shaw, Adam L. and Chen, Zhuo and Choi, Joonhee and Mark, Daniel K. and Scholl, Pascal and Finkelstein, Ran and Elben, Andreas and Choi, Soonwon and Endres, Manuel},
	date = {2024/04/01},
	doi = {10.1038/s41586-024-07173-x},
	id = {Shaw2024},
	isbn = {1476-4687},
	journal = {Nature},
	number = {8006},
	pages = {71--77},
	title = {Benchmarking highly entangled states on a 60-atom analogue quantum simulator},
	url = {https://doi.org/10.1038/s41586-024-07173-x},
	volume = {628},
	year = {2024}}

@article{Labuhn_2016,
	author = {Labuhn, Henning and Barredo, Daniel and Ravets, Sylvain and de L{\'e}s{\'e}leuc, Sylvain and Macr{\`\i}, Tommaso and Lahaye, Thierry and Browaeys, Antoine},
	date = {2016/06/01},
	doi = {10.1038/nature18274},
	id = {Labuhn2016},
	isbn = {1476-4687},
	journal = {Nature},
	number = {7609},
	pages = {667--670},
	title = {{Tunable two-dimensional arrays of single Rydberg atoms for realizing quantum Ising models}},
	url = {https://doi.org/10.1038/nature18274},
	volume = {534},
	year = {2016}}

@article{Leseleuc_2019,
	author = {Sylvain de L{\'e}s{\'e}leuc and Vincent Lienhard and Pascal Scholl and Daniel Barredo and Sebastian Weber and Nicolai Lang and Hans Peter B{\"u}chler and Thierry Lahaye and Antoine Browaeys},
	doi = {10.1126/science.aav9105},
	journal = {Science},
	number = {6455},
	pages = {775-780},
	title = {{Observation of a symmetry-protected topological phase of interacting bosons with Rydberg atoms}},
	url = {https://www.science.org/doi/abs/10.1126/science.aav9105},
	volume = {365},
	year = {2019}}

@article{Browaeys_2020,
	author = {Browaeys, Antoine and Lahaye, Thierry},
	date = {2020/02/01},
	doi = {10.1038/s41567-019-0733-z},
	id = {Browaeys2020},
	isbn = {1745-2481},
	journal = {Nature Physics},
	number = {2},
	pages = {132--142},
	title = {{Many-body physics with individually controlled Rydberg atoms}},
	url = {https://doi.org/10.1038/s41567-019-0733-z},
	volume = {16},
	year = {2020}}

@article{Kaufman_2021,
	author = {Kaufman, Adam M. and Ni, Kang-Kuen},
	date = {2021/12/01},
	doi = {10.1038/s41567-021-01357-2},
	id = {Kaufman2021},
	isbn = {1745-2481},
	journal = {Nature Physics},
	number = {12},
	pages = {1324--1333},
	title = {{Quantum science with optical tweezer arrays of ultracold atoms and molecules}},
	url = {https://doi.org/10.1038/s41567-021-01357-2},
	volume = {17},
	year = {2021}}

@article{Moessner_2000,
  title = {{Two-Dimensional Periodic Frustrated Ising Models in a Transverse Field}},
  author = {Moessner, R. and Sondhi, S. L. and Chandra, P.},
  journal = {Phys. Rev. Lett.},
  volume = {84},
  issue = {19},
  pages = {4457--4460},
  numpages = {0},
  year = {2000},
  month = {May},
  publisher = {American Physical Society},
  doi = {10.1103/PhysRevLett.84.4457},
  url = {https://link.aps.org/doi/10.1103/PhysRevLett.84.4457}
}

@article{Humeniuk_2016,
  title = {{Quantum Monte Carlo study of long-range transverse-field Ising models on the triangular lattice}},
  author = {Humeniuk, Stephan},
  journal = {Phys. Rev. B},
  volume = {93},
  issue = {10},
  pages = {104412},
  numpages = {13},
  year = {2016},
  month = {Mar},
  publisher = {American Physical Society},
  doi = {10.1103/PhysRevB.93.104412},
  url = {https://link.aps.org/doi/10.1103/PhysRevB.93.104412}
}

@article{Koziol_2019,
  title = {{Quantum criticality of the transverse-field Ising model with long-range interactions on triangular-lattice cylinders}},
  author = {Koziol, Jan and Fey, Sebastian and Kapfer, Sebastian C. and Schmidt, Kai Phillip},
  journal = {Phys. Rev. B},
  volume = {100},
  issue = {14},
  pages = {144411},
  numpages = {11},
  year = {2019},
  month = {Oct},
  publisher = {American Physical Society},
  doi = {10.1103/PhysRevB.100.144411},
  url = {https://link.aps.org/doi/10.1103/PhysRevB.100.144411}
}

@book{Sachdev_2023, place={Cambridge}, title={{Quantum Phases of Matter}}, publisher={Cambridge University Press}, author={Sachdev, Subir}, year={2023}}

@book{Wen_2007,
	title = {Quantum {Field} {Theory} of {Many}-{Body} {Systems}: {From} the {Origin} of {Sound} to an {Origin} of {Light} and {Electrons}},
	isbn = {978-0-19-922725-9},
	publisher = {Oxford University Press},
	author = {Wen, Xiao-Gang},
	month = sep,
	year = {2007},
	doi = {10.1093/acprof:oso/9780199227259.001.0001}
}

@article{Altman_2021,
  title = {{Quantum Simulators: Architectures and Opportunities}},
  author = {Altman, Ehud and Brown, Kenneth R. and Carleo, Giuseppe and Carr, Lincoln D. and Demler, Eugene and Chin, Cheng and DeMarco, Brian and Economou, Sophia E. and Eriksson, Mark A. and Fu, Kai-Mei C. and Greiner, Markus and Hazzard, Kaden R.A. and Hulet, Randall G. and Koll\'ar, Alicia J. and Lev, Benjamin L. and Lukin, Mikhail D. and Ma, Ruichao and Mi, Xiao and Misra, Shashank and Monroe, Christopher and Murch, Kater and Nazario, Zaira and Ni, Kang-Kuen and Potter, Andrew C. and Roushan, Pedram and Saffman, Mark and Schleier-Smith, Monika and Siddiqi, Irfan and Simmonds, Raymond and Singh, Meenakshi and Spielman, I.B. and Temme, Kristan and Weiss, David S. and Vu\ifmmode \check{c}\else \v{c}\fi{}kovi\ifmmode \acute{c}\else \'{c}\fi{}, Jelena and Vuleti\ifmmode \acute{c}\else \'{c}\fi{}, Vladan and Ye, Jun and Zwierlein, Martin},
  journal = {PRX Quantum},
  volume = {2},
  issue = {1},
  pages = {017003},
  numpages = {19},
  year = {2021},
  month = {Feb},
  publisher = {American Physical Society},
  doi = {10.1103/PRXQuantum.2.017003},
  url = {https://link.aps.org/doi/10.1103/PRXQuantum.2.017003}
}

@article{Schollwock_2011,
title = {{The density-matrix renormalization group in the age of matrix product states}},
journal = {Annals of Physics},
volume = {326},
number = {1},
pages = {96-192},
year = {2011},
issn = {0003-4916},
doi = {https://doi.org/10.1016/j.aop.2010.09.012},
author = {Ulrich Schollwöck}
}

@Article{tenpy2024,
    title={{Tensor network Python (TeNPy) version 1}},
    author={Johannes Hauschild and Jakob Unfried and Sajant Anand and Bartholomew Andrews and Marcus Bintz and Umberto Borla and Stefan Divic and Markus Drescher and Jan Geiger and Martin Hefel and Kévin Hémery and Wilhelm Kadow and Jack Kemp and Nico Kirchner and Vincent S. Liu and Gunnar Möller and Daniel Parker and Michael Rader and Anton Romen and Samuel Scalet and Leon Schoonderwoerd and Maximilian Schulz and Tomohiro Soejima and Philipp Thoma and Yantao Wu and Philip Zechmann and Ludwig Zweng and Roger S. K. Mong and Michael P. Zaletel and Frank Pollmann},
    journal={SciPost Phys. Codebases},
    pages={41},
    year={2024},
    publisher={SciPost},
    doi={10.21468/SciPostPhysCodeb.41},
    url={https://scipost.org/10.21468/SciPostPhysCodeb.41},
}

@article{SM,
      title = {},
      author = {},
      journal = {See Supplemental Material for further details on the effective field theory and the DMRG numerics, including a discussion on the interaction tails, the role of the driving field, and realistic experimental parameters for a possible state preparation.},
      volume = {},
      issue = {},
      pages = {},
      numpages = {},
      year = {},
      month = {},
      publisher = {},
      doi = {},
      url = {}
    }

@book{DiFrancesco_CFT_book,
      author        = "Di Francesco, Philippe and Mathieu, Pierre and Sénéchal,
                       David",
      title         = "{Conformal field theory}",
      publisher     = "Springer",
      address       = "New York, NY",
      year          = "1997"
}

@article{Matsuo_2006,
    doi = {10.1088/0305-4470/39/12/006},
    url = {https://dx.doi.org/10.1088/0305-4470/39/12/006},
    year = {2006},
    month = {mar},
    publisher = {},
    volume = {39},
    number = {12},
    pages = {2953},
    author = {Matsuo, Haruhiko and Nomura, Kiyohide},
    title = {{Berezinskii–Kosterlitz–Thouless transitions in the six-state clock model}},
    journal = {Journal of Physics A: Mathematical and General}
    }

@article{Zhou_2025,
    doi = {10.1088/0256-307X/42/5/053705},
    url = {https://dx.doi.org/10.1088/0256-307X/42/5/053705},
    year = {2025},
    month = {may},
    publisher = {Chinese Physical Society and IOP Publishing Ltd},
    volume = {42},
    number = {5},
    pages = {053705},
    author = {Zhou, Zheng and Yan, Zheng and Liu, Changle and Chen, Yan and Zhang, Xue-Feng},
    title = {{Quantum Simulation of Two-Dimensional U(1) Gauge Theory in Rydberg and Rydberg-Dressed Atom Arrays}},
    journal = {Chinese Physics Letters}
    }

@article{Li_2020,
  title = {Critical properties of the two-dimensional $q$-state clock model},
  author = {Li, Zi-Qian and Yang, Li-Ping and Xie, Z. Y. and Tu, Hong-Hao and Liao, Hai-Jun and Xiang, T.},
  journal = {Phys. Rev. E},
  volume = {101},
  issue = {6},
  pages = {060105},
  numpages = {5},
  year = {2020},
  month = {Jun},
  publisher = {American Physical Society},
  doi = {10.1103/PhysRevE.101.060105}
}

@article{Delfino_1998,
	author = {G. Delfino and G. Mussardo},
	doi = {https://doi.org/10.1016/S0550-3213(98)00063-7},
	issn = {0550-3213},
	journal = {Nuclear Physics B},
	number = {3},
	pages = {675-703},
	title = {{Non-integrable aspects of the multi-frequency sine-Gordon model}},
	url = {https://www.sciencedirect.com/science/article/pii/S0550321398000637},
	volume = {516},
	year = {1998}}

@article{Prakash_2025,
  title = {{Classical Origins of Landau-Incompatible Transitions}},
  author = {Prakash, Abhishodh and Jones, Nick G.},
  journal = {Phys. Rev. Lett.},
  volume = {134},
  issue = {9},
  pages = {097103},
  numpages = {9},
  year = {2025},
  month = {Mar},
  publisher = {American Physical Society},
  doi = {10.1103/PhysRevLett.134.097103},
  url = {https://link.aps.org/doi/10.1103/PhysRevLett.134.097103}
}

@article{Huang_2020,
  title = {{Kibble-Zurek mechanism for a one-dimensional incarnation of a deconfined quantum critical point}},
  author = {Huang, Rui-Zhen and Yin, Shuai},
  journal = {Phys. Rev. Res.},
  volume = {2},
  issue = {2},
  pages = {023175},
  numpages = {7},
  year = {2020},
  month = {May},
  publisher = {American Physical Society},
  doi = {10.1103/PhysRevResearch.2.023175},
  url = {https://link.aps.org/doi/10.1103/PhysRevResearch.2.023175}
}

@article{Shu_2025,
	title = {Equilibration of topological defects near the deconfined quantum multicritical point},
	volume = {16},
	issn = {2041-1723},
	url = {https://doi.org/10.1038/s41467-025-58477-z},
	doi = {10.1038/s41467-025-58477-z},
	pages = {3402},
	number = {1},
	journal = {Nature Communications},
	author = {Shu, Yu-Rong and Jian, Shao-Kai and Sandvik, Anders W. and Yin, Shuai},
	date = {2025-04-10},
}

@dataset{data,
  author       = {Lisa Bombieri and
                  Zache, Torsten Victor and
                  Calliari, Gabriele and
                  Mikhail, Lukin and
                  Pichler, Hannes and
                  González-Cuadra, Daniel},
  title        = {Deconfined quantum criticality on a triangular
                   Rydberg array
                  },
  year         = 2025,
  publisher    = {Zenodo},
  doi          = {10.5281/zenodo.16599117},
  url          = {https://doi.org/10.5281/zenodo.16599117},
}
\let\addcontentsline\oldaddcontentsline

\clearpage
\onecolumngrid
\begin{center}
    \textbf{\Large Supplemental Material to 
``Deconfined quantum criticality on a triangular Rydberg array''}
\end{center}

\normalsize

\setcounter{equation}{0}
\setcounter{figure}{0}
\setcounter{table}{0}
\makeatletter
\renewcommand{\theequation}{S\arabic{equation}}
\renewcommand{\thefigure}{S\arabic{figure}}
\setlength\tabcolsep{10pt}
\setcounter{secnumdepth}{3}

\newcommand\numberthis{\addtocounter{equation}{1}\tag{\theequation}}
\newcommand{\insertimage}[1]{\includegraphics[valign=c,width=0.04\columnwidth]{#1}}
\onecolumngrid
In this Supplementary Material, we provide additional details supporting the analytical and numerical results presented in the main text. In App.~\ref{sec:SM_appA}, we expand on the field-theoretical analysis. In App.~\ref{sec:SM_numerics}, we detail the numerical methods, including the data used to extract the critical exponents shown in Fig.~2(f), a finite-size scaling analysis confirming the dependence of the critical exponents on the cylinder circumference, an analysis of the statistics of the U($1$)-symmetric state for cylinders of different circumferences, and a discussion on the truncation of the interaction tails. In App.~\ref{sec:App_open_systems}, we consider the open system in Fig.~3(c)-(f) with interactions up to the fifth-nearest neighbor and demonstrate how the staggered magnetization serves as an experimentally accessible observable to distinguish between the ordered
phases and the critical region. In App.~\ref{sec:SM_vary_Omega}, we examine the dependence on $\Omega/U$ for the cylinder with the smallest circumference considered in the main text and show that the field-theory description---and thus the DQCP---remains valid for $\Omega/U\lesssim 0.43$, with a Luttinger parameter $K$ that depends on $\Omega/U$. Finally, in App.~\ref{App:exp_paramters}, we provide realistic experimental parameters for
a possible state preparation of the configuration shown in
Fig.~3(c).

\setcounter{equation}{0}
\setcounter{figure}{0}
\setcounter{table}{0}
\renewcommand{\theequation}{S\arabic{equation}}
\renewcommand{\thefigure}{S\arabic{figure}}
\setlength\tabcolsep{10pt}
\setcounter{secnumdepth}{3}

\tableofcontents
\vspace{4mm}
\twocolumngrid
\section{Details on the field-theoretical description}\label{sec:SM_appA}
Here, we provide further details on the field theory discussed in the main text. Our analysis begins with the formulation of a low-energy 2D effective field theory for the complex staggered magnetization $m=\rho e^{i\phi}$ [Eq.~$(2)$]. From this, we then derive an effective field theory governing its phase $\phi$. We then discuss it in the 1D limit, corresponding to small circumferences. Finally, we present analytical predictions for the marginal probability distributions $P(\phi)$ and $P(\rho)$. 

\subsection{Mapping}
To formulate the 2D field theory, we begin by defining how the microscopic lattice model is mapped onto a continuum description.
We consider blocks of size $3 \times 3$, corresponding to the unit cells of the considered Rydberg model [Eq.~$(1)$]. We label each unit cell by ${\bf u} = (u_x, u_y) \in \mathbb{Z}^2$, e.g., indicating the position of its `center' ${\bf r}_{\bf u} = \left(u_x + \tfrac{1}{2}\right) {\bf A}_x + \left(u_y + \tfrac{1}{2}\right) {\bf A}_y$, where ${\bf A}_x = 3\,{\bf a}_x$ and ${\bf A}_y = 3\,{\bf a}_y$ are the basis vectors of the effective lattice with lattice spacing $3a$. Here, ${\bf a}_x$ and ${\bf a}_y$ are the primitive vectors, and $a$ is the lattice spacing of the underlying triangular lattice. 

On this coarse-grained lattice, we define a local observable on each unit cell given by the staggered magnetization $m({\bf u}) = \sum_{{\bf j} \in {\bf u}} n_{\bf j}\, e^{i\,{\bf Q} \cdot x_{\bf j}}$,
where the sum runs over all sites ${\bf j}$ within the $3 \times 3$ unit cell centered at ${\bf r}_{\bf u}$.
In the `continuum limit', where $3a\to0$, we associate with each discrete site ${\bf u} \in \mathbb{Z}^2$ a continuous position ${\bf r}=(x,y) \in \mathbb{R}^2$ and to the observable defined on the discrete lattice $m({\bf u})$ a smooth field $\psi({\bf r})$ such that $\psi({\bf r}) = \lim_{a \to 0} m({\bf u})$.
Throughout the numerics, we set $a = 1$ and consider the limit of many unit cells to approximate the continuum behavior.

\subsection{Field theory in 2D}\label{secSM:field_th_2D}
Here, we formulate a continuous 2D field-theoretic description of the system in terms of the complex scalar field $\psi({\bf r})$. 
Guided by the symmetries of the two ordered phases and the emergence of a U($1$) symmetry at criticality of the Rydberg model [Fig.~1], we propose the following 2D effective theory, described by the Euclidean-space path integral $Z=\int D[\psi, \psi^*] \, e^{-S[\psi, \psi^*]}$, with associated action
\begin{widetext}
\begin{equation}
    S[\psi, \psi^*] = \int d\tau \, dx  \int_0^{l_y} dy  \left[ \frac{|\partial_\mu \psi|^2}{2\pi K} - \mu |\psi|^2 + \lambda |\psi|^4 + \frac{g_3}{2} \left(\psi^3 + \psi^{*3} \right) + \frac{g_6}{2} \left( \psi^6 + \psi^{*6} \right) \right].
    \label{eq:field_th_full}
\end{equation}
\end{widetext}
Here, $|\partial_\mu \psi|^2=|\partial_\tau \psi|^2+|\nabla \psi|^2$, $\tau$ parametrizes the imaginary time, and the $y$-direction is compact: $\psi(\tau,x,l_y)=\psi(\tau,x,0)$. Moreover, $g_3$ and $g_6$ are the coupling constants of the $\mathbb{Z}_3$ and the $\mathbb{Z}_6$ symmetry-breaking perturbations, respectively. To describe the transition between the ordered phases at $1/3$- and the $2/3$-filling, we require $g_3$ to change sign across the transition and vanish at criticality.  Due to the symmetry of the action under the exchange of imaginary time $\tau$ and the compact spatial coordinate $y$, this theory on a spatial cylinder can be interpreted as describing an isotropic system at finite temperature $k_{\rm B} T=1/\beta=1/l_y$.

Finally, we note that higher-order interaction terms in the action become irrelevant under renormalization group (RG) flow at low energies due to their negative mass dimension. Indeed, assuming $K$ is dimensionless, i.e., $[K]=0$, the field $\psi$ then has mass dimension $[\psi] = 1/2$, and the coupling constant $g_n$ associated with the perturbation $\psi^n + \psi^{*n}$ has mass dimension $[g_n]=3-n/2$. 

We now switch to polar coordinates by introducing real fields $\rho$ and $\phi$ via the parametrization $\psi=\rho e^{i\phi}$. The propagator becomes $Z=\int D[\rho, \phi] \, \, e^{-S[\rho,\phi]}$, where the measure is $D[\rho,\phi]= \rho\, d\rho d\phi$, and the action is given by
\begin{widetext}
\begin{equation}
  S[\rho, \phi] = \int d \tau \, dx  \int_0^{l_y} dy  \left[ \frac{(\partial_\mu \rho)^2}{2\pi K} - \mu \rho^2 + \lambda \rho^4 +  \frac{\rho^2}{2\pi K}(\partial_\mu \phi)^2  + g_3 \rho^3\cos{(3\phi)} + g_6 \rho^6 \cos{(6\phi)} \right],
  \label{eq:field_th_full_rho_phi}
\end{equation}
\end{widetext}
where we have used the identity $|\partial_\mu \psi|^2=(\partial_\mu  \rho)^2 + \rho^2 (\partial_\mu \phi)^2$.
To try to solve this model, we aim to factorize the propagator as $Z=Z_\rho Z_\phi$, where $Z_\rho=\int D[\rho] \, e^{-S[\rho]}$ and $Z_\phi=\int D[\phi] \, e^{-S[\phi]}$. This requires that the action decomposes into two independent parts: $S[\rho, \phi]=S[\rho]+S[\phi]$. To achieve this, we need to make the approximations and assumptions detailed in the following.

\subsection{Effective field theory for the phase field: from 2D to 1D}
First, we neglect fluctuations of the amplitude field $\rho$ and assume it is fixed at the constant value $\rho_0$ that minimizes the action. This corresponds to a saddle-point approximation, where $\rho_0$ satisfies
\begin{equation}
    \left. \frac{\partial S}{\partial \rho} \right|_{\rho = \rho_0} =  2\rho_0 (2\lambda \rho_0^2-\mu)=0,
\end{equation}
yielding to
\begin{equation}
    \rho_0=0 \:\:\ \text{for} \:\:\ \mu<0 \:\:\ \text{and} \:\:\ \rho_0=\sqrt{\frac{\mu} {2\lambda}}  \:\:\ \text{for} \:\:\ \mu>0.
\end{equation}

Second, we assume the system is in the phase with $\mu>0$. 
Substituting $\rho=\rho_0$ into Eq.~\eqref{eq:field_th_full_rho_phi}, we obtain the following effective action for the phase field $\phi$: 
\begin{widetext}    
\begin{equation}
    S[\phi]  = \int d \tau  dx \int_0^{l_y}  dy   \left[\frac{\rho_0^2}{2\pi K}(\partial_\mu \phi)^2  + g_3 \rho_0^3 \cos{(3\phi)} + g_6 \rho_0^6 \cos{(6\phi)}\right].
    \label{eq:Sphi_2D}
\end{equation}
\end{widetext}

\subsubsection{Effective 1D model}
We now perform a dimensional reduction of the 2D field theory in Eq.~\eqref{eq:Sphi_2D} to an effective 1D theory.  Specifically, for a small cylinder circumference $l_y$, we assume the system is translationally invariant along the $y$-direction [$\phi(\tau,x,y)\to \phi(\tau,x)$], and we obtain the effective 1D action
\begin{equation}
    S[\phi]=\int d\tau \, dx \left[\frac{(\partial_\mu \phi)^2 }{2\pi K'} + g'_3 \cos{(3\phi)} + g'_6 \cos{(6\phi)} \right],
    \label{eq:Sphi_g3g6}
\end{equation}
with parameters
\begin{equation}
    K' = K/(\rho_0^2l_y), \:\:\ g'_3 = g_3 l_y  \rho_0^3, \:\:\ \text{and} \:\:\ g'_6= g_6 l_y \rho_0^6.
    \label{eq:effth1D_paramteres}
\end{equation}
In particular, we find that $K'$ decreases as the cylinder circumference $l_y$ increases, and therefore increases with temperature, i.e., $K'\sim 1/l_y \sim T$ (see Sec.~\ref{secSM:field_th_2D}). Thus, analyzing the phase diagram of this model for varying $K'$ corresponds to analyzing the phase diagram of the original Rydberg model as a function of temperature.

While Eq.~\eqref{eq:Sphi_g3g6} captures the low-temperature physics, to describe the high-temperature behavior we include in the following the term $g'_H\cos{(2\pi\Theta)}$, in analogy to the $\mathbb{Z}_q$ deformed sine-Gordon model~\cite{Matsuo_2006}:
\begin{align}
    S[\phi] = \int d\tau \, dx \Bigg[ \frac{(\partial_\mu \phi)^2}{2\pi K'} 
    + g'_3 \cos(3\phi) + g'_6 \cos(6\phi) \nonumber \\
    + g'_H \cos\left(2\pi\Theta\right) \Bigg].
    \label{eq:Sphi_g3g6_gH}
\end{align} 
Here, the field $\Theta$ is mutually dual to $\phi$, satisfying the relation $\partial_x \phi = -\pi K'\partial_\tau\Theta$ and $\partial_\tau \phi = \pi K' \partial_x \Theta$. We note that this model has also been recently discussed in Ref.~\cite{Prakash_2025}.

\subsubsection{Phase diagram}
\begin{figure}
    \centering
    \includegraphics[width=\columnwidth]{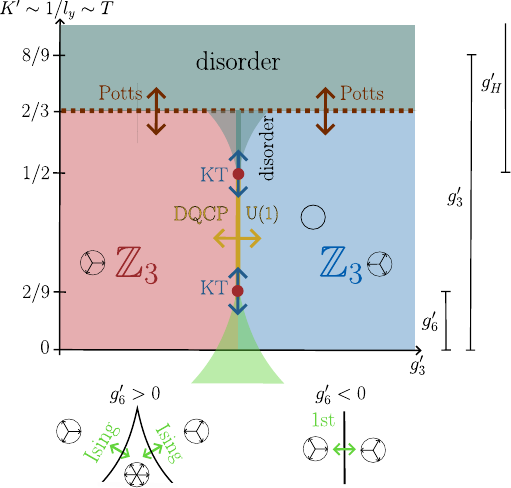}
    \caption{Phase diagram and phase transitions for the model in Eq.~\eqref{eq:Sphi_g3g6_gH}. }
    \label{fig:effth_g3g6}
\end{figure}
The phase diagram of the effective 1D model in Eq.~\eqref{eq:Sphi_g3g6_gH} can be understood by analyzing the renormalization group (RG) flow of the perturbative terms~\cite{DiFrancesco_CFT_book}. The scaling dimension of a general operator $O_{m,n} = e^{in\phi} e^{im\Theta}$ is given by~\cite{Matsuo_2006}
\begin{equation}
    x_{m,n} = \frac{1}{4}\left(\frac{m^2}{K'} + n^2 K'\right).
\label{eq:scaling_dim_x3}
\end{equation}
In particular, the scaling dimensions of $\cos{(3\phi)}$,  $\cos{(6\phi)}$, and  $\cos{(2\pi\Theta)}$ are
\begin{equation}
    x_3 = \frac{9}{4}K', \:\:\  x_6 = 9K' , \:\:\ \text{and} \:\:\ x_{\Theta} = \frac{1}{K'}.
\end{equation}
The relevance or irrelevance of each term depends on whether its scaling dimension is less than or greater than $2$, respectively. This allows us to distinguish different regimes as a function of $K'$ (interpreted as temperature), as illustrated in the phase diagram in Fig.~\ref{fig:effth_g3g6}. 

Let us first consider the case $g'_3\neq 0$ and the region $K'>2/9$. Here, the $\mathbb{Z}_6$ symmetry breaking term $\cos(6 \phi)$ becomes irrelevant, effectively reducing the theory in Eq.~\eqref{eq:Sphi_g3g6_gH} to a 3-state clock model~\cite{Li_2020}.
Upon varying $K'$ (equivalent to $T$), we encounter a phase transition between a low-temperature ordered phase, in which the $\mathbb{Z}_3$ symmetry is spontaneously broken, and a high-temperature disordered phase, characterized by exponentially decaying correlations. This transition belongs to the universality class of the Potts model. The critical point is determined by the so-called `self-dual' condition $x_3=x_\Theta$, which is satisfied for $K'=2/3$. These findings are consistent with the transition between the ordered phase at $1/3$ (or at $2/3$)-filling  and the high-temperature disordered phase of the Rydberg model described in Ref.~\cite{Guo_2023}.

Next, we fix $K'$ within the $\mathbb{Z}_3$-ordered phase, i.e., $2/9<K'<2/3$. In this range, upon tuning $g'_3$ from negative to positive the system undergoes a continuous phase transition---a DQCP---between two distinct $\mathbb{Z}_3$ symmetry-broken phases. At the critical point, $g'_3=0$, the model reduces to a LL theory and exhibits power-law correlations and emerging $U(1)$ symmetry. This transition corresponds to the one analyzed in the main text between the ordered phases at $1/3$ and $2/3$ filling.

We now turn to the case $g'_3=0$ for any value of $K'$. Along this line, the theory in Eq.~\eqref{eq:Sphi_g3g6_gH} reduces to a 6-state clock model~\cite{Li_2020}. Unlike the 3-state clock model, this model presents three phases, separated by two KT-transitions: a low-temperature $\mathbb{Z}_6$-ordered phase for $K'<K'_1=2/9$ (where $\cos(6\phi)$ is a relevant perturbation), a high-temperature disordered phase for $K'>K'_2=1/2$ (where $\cos(2\pi\Theta)$ is a relevant perturbation), and an intermediate phase for $K_1'< K < K_2'$ (where none of the cosine is relevant). 
The latter is a critical phase with power-law decaying correlations (quasi-long-range order) and emergent $U(1)$ symmetry. This regime corresponds to the line of DQCPs depicted in the phase diagram in Fig.~\ref{fig:effth_g3g6}. The high-temperature transition agrees with the Kosterlitz-Thouless (KT) transition observed in Ref.~\cite{Guo_2023} and correctly predicts that this transition occurs at a lower temperature than the Potts transitions.

Finally, we discuss the transition between the two $\mathbb{Z}_3$-ordered phases for $K'<2/9$. In this regime, for $g_3'\neq 0$ both $\cos{(3\phi)}$ and $\cos{(6\phi)}$ are relevant, and the model reduces to a double-frequency sine-Gordon model~\cite{Delfino_1998}. The nature of the transition between the two $\mathbb{Z}_3$ phases for $|g'_3| \gg |g'_6|$ (with $g'_3>0$ or $g'_3<0$) depends on the sign of $g'_6$. It is a first-order transition for $g'_6<0$, while an intermediate phase with $\mathbb{Z}_6$ symmetry arises for $g'_6>0$. The phase transitions at the boundaries of this intermediate phase belong to the Ising universality class~\cite{Prakash_2025}. 
We note that in this regime of small $K'$, the dimensional reduction may break down, and the effective 1D field theory [Eq.~\eqref{eq:Sphi_g3g6}] may no longer capture the full 2D physics [Eq.~\eqref{eq:field_th_full_rho_phi}]. Indeed, our numerical simulations do not show clear evidence for either a first-order transition or an intermediate phase, as further discussed in App.~\ref{sec:SM_numerics}.

\subsubsection{Critical exponents}
Let us focus on the region $2/9<K'<8/9$, where $\cos{(3\phi)}$ is relevant and $\cos{(6\phi)}$ is irrelevant. Here, we can express the critical exponents only in terms of the scaling dimension $x_3$ of the perturbation $\cos{(3\phi)}$ in Eq.~\eqref{eq:Sphi_g3g6}. At the critical point, where $g'_3=0$, the two-point correlation function for the operator $\psi_n=\rho_0e^{i n \phi(x)}$ between two points $x_1$ and $x_2$ at a distance $r=|x_1-x_2|$ decays as a power-law~\cite{DiFrancesco_CFT_book}:
\begin{equation}
     C_n(r) =\rho^2_0 \langle e^{i n \phi(r)} e^{-i n \phi(0)} \rangle \sim r^{-2x_n}.
     \label{eq:correlation_funct_criticality}
\end{equation}  
For the order parameter $n=3$, the exponent $\eta = 2x_3$ is the anomalous dimension. 

A finite $g'_3$ drives the system away from the critical point, and the scaling of the correlation length $\xi$ and the order parameter $\langle \cos(3\phi) \rangle$ close to the transition are~\cite{DiFrancesco_CFT_book}:
\begin{align}
   & \xi \sim \left|g'_3\right|^{-\nu}, \:\:\ \text{with} \:\:\ \nu = \frac{1}{2 - x_3}, \label{eq:exponent_nu} \\
   & |\langle \cos(3\phi) \rangle | \sim \left| g'_3 \right|^\beta, \:\:\ \text{with} \:\:\ \beta = \frac{\eta \nu}{2} = \frac{x_3}{2 - x_3} \label{eq:exponent_beta}.
\end{align}
Notably, these exponents depend on the cylinder circumference via Eq.~\eqref{eq:effth1D_paramteres} and Eq.~\eqref{eq:scaling_dim_x3}. Specifically, for $\beta/\nu$ and $1/\nu$, we have:
\begin{equation}
    \frac{\beta}{\nu} = x_3 = \frac{9 K}{4 \rho^2_0}\, \frac{1}{l_y}, \:\:\ 
    \frac{1}{\nu} = 2-x_3 =  2-  \frac{9 K}{4  \rho^2_0} \, \frac{1}{l_y}.
    \label{eq:critical_exponent_beta_nu_ly}
\end{equation}

\subsection{Prediction of the marginal probabilities distributions}
To predict the marginal probability distributions for the phase and amplitude fields, we return to the 2D field theory in Eq.~\eqref{eq:field_th_full_rho_phi}.
For simplicity, to derive a nontrivial probability distribution for $\rho$, we assume that the amplitude field $\rho$ decouples from the phase field $\phi$ and is described independently by the following action: 
\begin{equation}
    S[\rho] = \int d\tau\,  dx \int_0^{l_y} dy \left[ \frac{(\partial_\mu \rho)^2}{2\pi K} - \mu \rho^2 + \lambda \rho^4 \right].
\end{equation}

Together with the phase action $S[\phi]$ given in Eq.~\eqref{eq:Sphi_2D}, these decoupled models allow us to define marginal probability distributions for the respective fields. Specifically, we define:  
\begin{align}
    &\langle \phi \rangle = \frac{1}{Z_\phi} \int d\phi \, \phi \, e^{-S[\phi]} =  \int d\phi \, \phi \, P(\phi) , \nonumber \\
    &\langle \rho \rangle = \frac{1}{Z_\rho} \int d\rho\, \rho \, \rho e^{-S[\rho]} = \int d\rho \, \rho \, P(\rho) ,
    \label{eq:marginal_probab_def}
\end{align}
where $P(\phi)$ and $P(\rho)$ are the marginal probability distributions for the phase and amplitude fields, respectively.

In our numerical analysis, we sample the `total' staggered magnetization $m=\sum_{\bf u} m(\bf u)$. We can relate this quantity to the zero-mode component of the field $\psi(x,y)$, given by $\tilde{\psi}(0,0)=\sum_{x,y} \psi(x,y)$, whose marginal probability distribution we can predict under certain assumptions. 
Assuming that the fields $\rho$ and $\phi$ are constant in space-time (in different unit cells), the marginal probability distributions of $\psi$ coincide with those of $\tilde{\psi}$. Furthermore, the actions simplify to: $S[\phi] =  \mathrm{Vol.} \, V[\phi]$ and $S[\rho] =  \mathrm{Vol.} \, V[\rho]$, where $\mathrm{Vol.}$ indicates the volume factor coming from the integrals, and the effective potentials are
\begin{align}
    &V[\phi] =  g_3 \rho^3_0 \cos{(3\phi)} + g_6 \rho^6_0 \cos{(6\phi)}, \\
    & V[\rho] =  -\mu \rho^2 + \lambda \rho^4 .
    \label{eq:potentials_Vphirho}
\end{align}
From these potentials, using Eq.~\eqref{eq:marginal_probab_def}, we obtain
\begin{align}
    P(\phi) = \frac{e^{-{\rm Vol.} \, V[\phi]}}{\int D[\phi] \, e^{-{\rm Vol.} \, V[\phi]}}, \label{eq:marginal_probab_phi} \\
    P(\rho) = \frac{\rho e^{-{\rm Vol.} \, V[\rho]}}{\int D[\rho] \, \rho e^{-{\rm Vol.} \, V[\rho]} }.
    \label{eq:marginal_probab_rho}
\end{align}
These expressions provide analytic predictions for the form of the marginal distributions that we compare to the numerical data in our work.

\section{Further details on the numerics on cylinders}\label{sec:SM_numerics}
\begin{figure*}
    \centering
    \includegraphics[width=\textwidth]{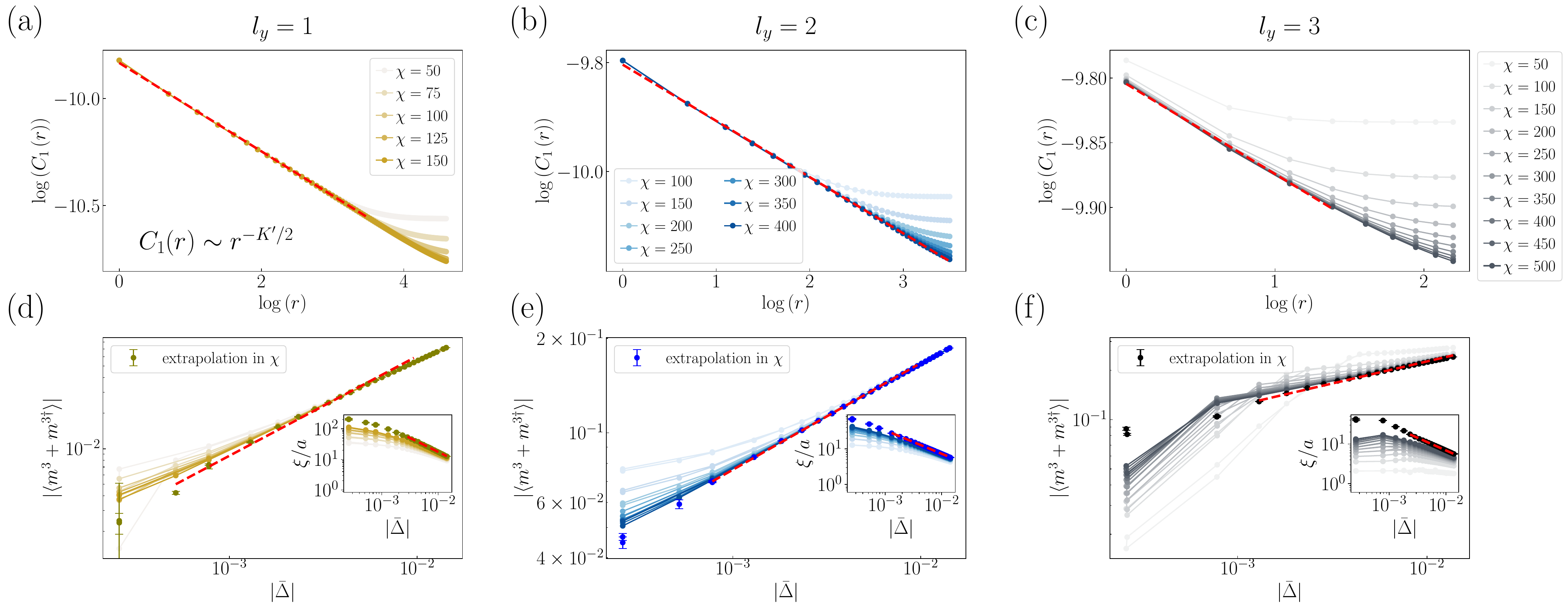}
    \caption{Analysis of infinitely long cylinders with (first column) $l_y=1$, (second column) $l_y=2$ and (third column) $l_y=3$ unit cells along ${\bf a}_y$. For each of them, we show for different bond dimensions $\chi$: (First row) Power-law decay of the two-point correlation $\langle m({\bf  r}) m(0)\rangle \approx C_1(r) =\rho^2_0\langle e^{i\phi(r)} e^{-i\phi(0)}\rangle$ at the transition point $(\Delta/U)_{\rm c}$, where the correlation length is maximal. The Luttinger parameter $K'$, shown in the inset of Fig.~2(d), is extracted from a linear log-log fit (red dashed line). (Second row) Evaluation of the critical exponent $\beta$ (main panels) and $\nu$ (insets), displayed in Fig.~2(f), from linear log-log fits of the order parameter and the correlation length as function of $\bar{\Delta}=\Delta/U-(\Delta/U)_{\rm c}$.}
    \label{fig:Fig_app_iMPS_cylinders}
\end{figure*}
Here, we provide further details on the numerical analysis. First, we discuss the numerical method employed. Second, we present the numerical data and the log-log fits used to determine the critical exponents shown in Fig.~2(f). Third, we perform a finite-size scaling analysis which confirms the dependence of the critical exponents on the cylinder circumference predicted by the effective 1D field theory [Eq.~\eqref{eq:critical_exponent_beta_nu_ly}]. Fourth, we analyze the statistics of the staggered magnetization on three finite cylinders with increasing circumference, by fitting the data to the analytical predictions for the marginal distributions given in Eqs.~\eqref{eq:marginal_probab_rho} and \eqref{eq:marginal_probab_phi}. This fitting procedure enables us to estimate how the ratio $|g_6/g_3|$ scales with the cylinder circumference. Finally, we include the interactions up to the fifth-nearest neighbor and show that the emergence of the U($1$) symmetry remains unaffected by the truncation at the third-nearest neighbors. 

\subsection{Computational methods}
Numerically, we represent the wavefunction using a matrix product state (MPS) ansatz and compute the ground state of the Hamiltonian via the density-matrix renormalization group (DMRG) algorithm, as implemented in the Python package TeNPy~\cite{tenpy2024}.
We map the lattice onto a one-dimensional MPS by wrapping it along the ${\bf a}_y$ direction. Our numerical results are converged in energy below error $10^{-6}$.
Our sweep strategy begins from a simple product state, progressively increasing the bond dimension. At each step, the converged MPS from the previous bond dimension is used as the initial state for the next. To avoid the algorithm becoming trapped in local minima, we include a mixer term during the early stages of the sweeps.

\subsection{Critical exponents derived from the numerics}
In Fig.~\ref{fig:Fig_app_iMPS_cylinders}, we show the linear fits used to extract the Luttinger parameter $K'$, the critical exponent $\beta$, and the critical exponent $\nu$ shown in Fig.~2, for infinitely long cylinders with $l_y=1,2,$ and $3$ unit cells along the $y$-axis. In particular, $K'$ follows from the decay of the two-point correlation function $\langle m(r) m(0)\rangle \approx C_1(r) = \rho_0^2 \langle e^{i\phi(r)} e^{-i\phi(0)}\rangle$, $\beta$ from the scaling of the order parameter $|\langle m^3 + m^{\dagger 3}\rangle| \sim |\bar{\Delta}|^{\beta}$, and  $\nu$ from the scaling of the correlation length $\xi \sim |\bar{\Delta}|^{-\nu}$. In the first case, we consider the decay of $C_1(r)$ for the largest considered bond dimension $\chi$ (as the results have converged). For the last two cases, we first extrapolate the observables as a function of  $\chi$ by performing a quadratic fit in $1/\chi$, and then apply a linear fit to the extrapolated values to extract the critical exponents. To estimate the error bars on these exponents, we repeat the fitting procedure over two different ranges: one over the full range shown in red, and another excluding one or two of the smallest $|\bar{\Delta}|$ values (for which the results are less converged in $\chi$). The difference between the resulting estimates is then taken as the uncertainty reported in Fig.~2(f).

Finally, for the largest circumference $l_y=3$, we find that the correlation length grows as $|\bar{\Delta}|$ is reduced, i.e., as the system approaches criticality, but drops at the very closest point we consider [see inset of Fig.~\ref{fig:Fig_app_iMPS_cylinders}(f)]. At first sight, this non-monotonic behavior might hint at an intermediate symmetry‑broken phase. However, up to $\chi=2000$, we find that the correlation length and entanglement entropy continue to scale as expected for a CFT with central charge $c\approx0.5$; they do not saturate as they would in a fully ordered phase. Moreover, at the second‑closest point to criticality we extract $c\approx1$ in disagreement with the Ising value $c=0.5$ that should appear at the boundaries of the intermediate ordered phase (Fig.~\ref{fig:effth_g3g6}). These observations suggest that the dip in correlation length is most likely a finite‑$\chi$ artifact rather than evidence of a genuine intermediate phase. However, further numerical investigation at larger bond dimensions is required to clarify the nature of this apparent dip in the correlation length.

\subsection{Finite size scaling}
\begin{figure}
    \centering
    \includegraphics[width=\columnwidth]{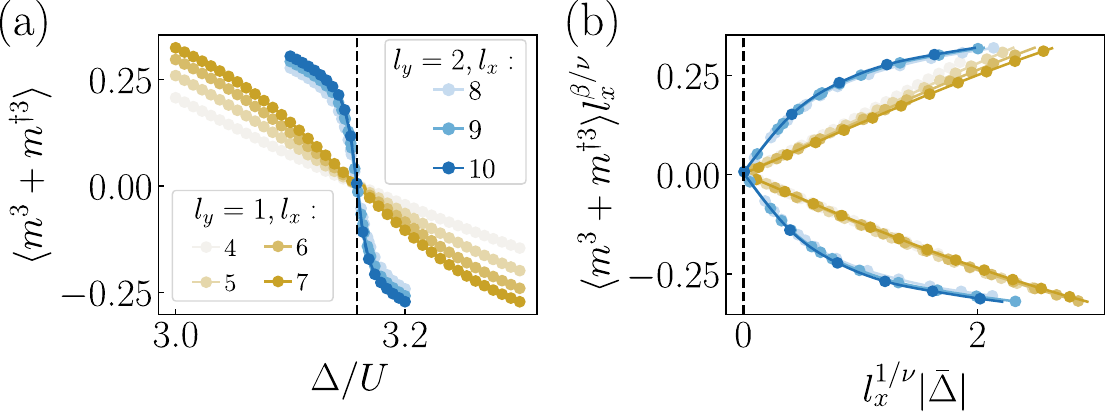}
    \caption{ Finite size scaling analysis on cylinders. 
(a) Order parameter $\langle m^3+m^{3 \dagger} \rangle$ as a function of $\Delta/U$ for cylinders with circumference (yellow) $l_y=1$ and (blue) $l_y=2$ unit cells along ${\bf a}_y$, and increasing lengths $l_x$ (unit cells along ${\bf a}_x$). The dashed vertical line is located at $\Delta/U=3.158$. (b) Finite-size scaling using the ansatz in Eq.~\eqref{eq:FSS}, with $\bar{\Delta}=\Delta/U-(\Delta/U)_{\rm c}$ and $(\Delta/U)_{\rm c}\approx3.158$. The exponents depend on $l_y$ according to Eq.~\eqref{eq:critical_exponent_beta_nu_ly}, with $ K/\rho_0^2 \approx 0.1$. Results obtained using a finite MPS with bond dimension $\chi=200$ for $l_y=1$, and $\chi=300$ for $l_y=2$.}
    \label{fig:FSS_finite_cylinders_nx_App}
\end{figure}

Here, we further numerically confirm the dependence of the critical exponents on the cylinder circumference predicted by the effective field theory [Eq.~\eqref{eq:critical_exponent_beta_nu_ly}] by performing a finite-size scaling (FSS) analysis in the cylinder length. In particular, we consider cylinders of increasing lengths $l_x= N_x/3$ ($l_x$ is the number of unit cells along ${\bf a}_x$) in the limit of $l_x\gg l_y$, i.e., $l_x>4l_y$.

For finite cylinders, the correlation length $\xi$ is bounded by the number of unit cells along the $x$-direction $l_x$, assuming that the data is converged in bond dimension $\chi$. Accordingly, we replace  $\xi \to l_x$ and identify the perturbation $g'_3$ in Eq.~\eqref{eq:critical_exponent_beta_nu_ly} with $\bar{\Delta}=\Delta/U-(\Delta/U)_{\rm c}$, arriving at the following scaling hypothesis:
\begin{equation}
    |\langle m^3 + m^{\dagger 3} \rangle| \sim l^{-\beta/\nu}_x \mathcal{F}\left(|\bar{\Delta}| l^{1/\nu}_x\right).
    \label{eq:FSS}
\end{equation}
where $\mathcal{F}$ is a universal scaling function, with $\mathcal{F}\to$ constant at the critical point $\bar{\Delta}=0$, and the exponents $\beta/\nu$ and $1/\nu$ depend on the cylinder circumference  $l_y$ as described in Eq.~\eqref{eq:critical_exponent_beta_nu_ly}. 

We first apply this scaling analysis to the smallest cylinder with $l_y=1$ [Fig.~\ref{fig:FSS_finite_cylinders_nx_App}], extracting the constant $K / \rho^2_0$ that ensures data collapse: $K / \rho^2_0 \approx 0.1$. (Note that this constant differs from the one extracted for infinite cylinders.) Using this value, we then verify that a consistent scaling collapse is obtained for the largest cylinder with $l_y=2$, confirming the dependence of the critical exponents on cylinder circumference. We note that cylinders with $l_y=3$ and large $l_x$ are beyond our current numerical reach.

\subsection{Sampling of the wavefunction on open cylinders and marginal probability distributions}
\begin{figure}
    \centering
    \includegraphics[width=\columnwidth]{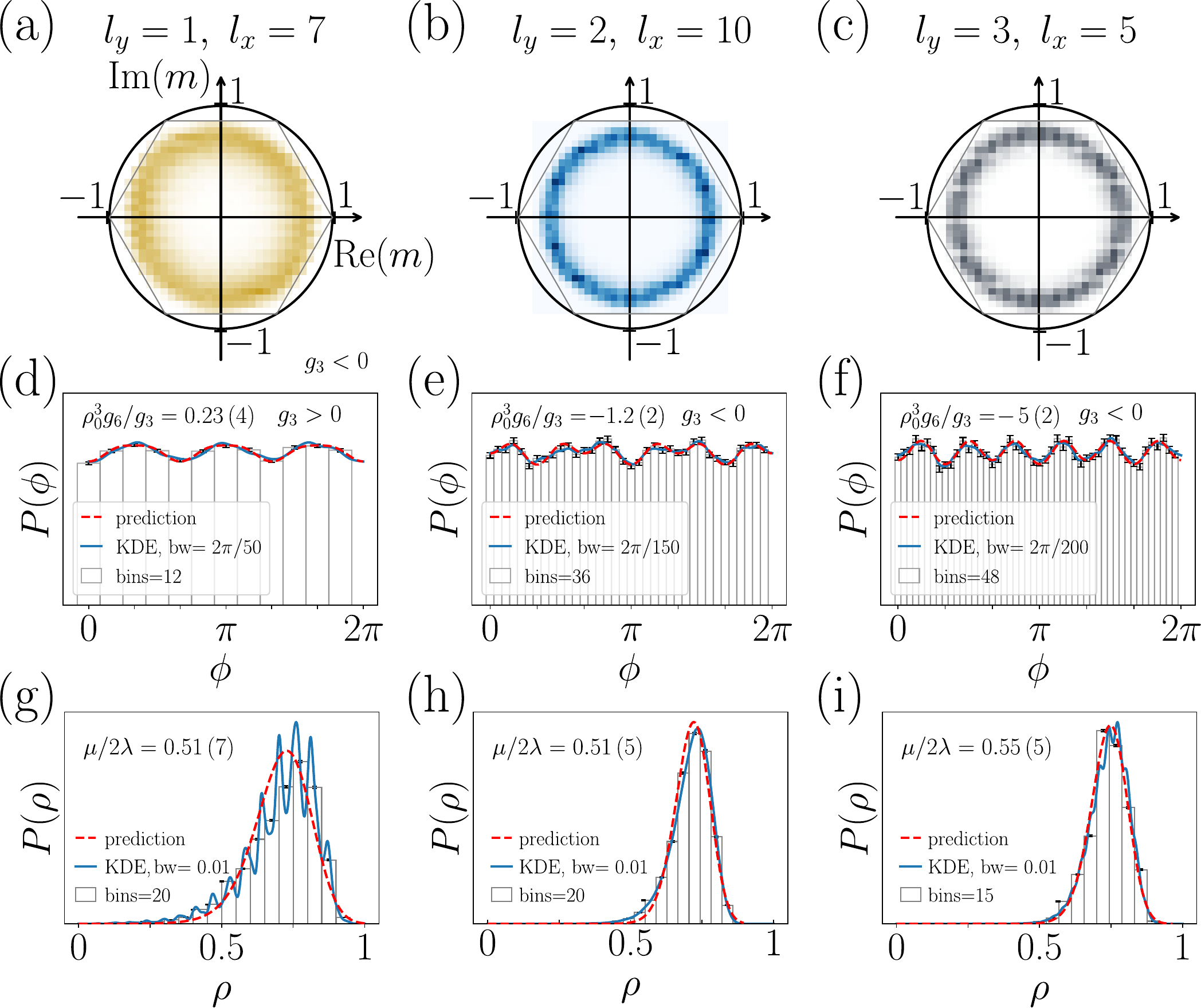}
    \caption{Statistics of the staggered magnetization $m$ [Eq.~$(2)$] for three open cylinders with increasing circumference $l_y$ at $\Delta/U \approx 3.1615$, $3.15789$, and $3.1584$, respectively. We consider: (First column) $l_y=1$, $l_x=7$ unit cells along ${\bf a}_y$ and ${\bf a}_x$, respectively; (Second column) $l_y=2$, $l_x=10$; and (Third column) $l_y=3$, $l_x=5$. (First row) Distribution of $m$ obtained from $10^5$ snapshots in the occupation basis, restricted to the bulk (excluding the first and last three rows of atoms along $\bf{a}_x$). (Second row) Corresponding angular distribution, fitted according to Eq.~\eqref{eq:marginal_probab_phi} (red dashed line). (Third row) Corresponding radial distribution, fitted according to Eq.~\eqref{eq:marginal_probab_rho} (red dashed line). Blue lines in (d)-(i) indicate a Gaussian kernel density estimate. Results obtained using a finite MPS with bond dimension $\chi=200$ for $l_y=1$, $\chi=250$ for $l_y=2$, and $\chi=500$ for $l_y=3$.}
    \label{fig:Sampling_open_cylinders_App}
\end{figure}
Here, we analyze the statistics of the staggered magnetization $m$ [Eq.~$(2)$] for open cylinders of increasing circumference $l_y$, near the transition point. We observe the emergence of a $U(1)$ symmetry and explain the marginal angular and radial probability distributions using the effective field theory discussed above. We note that the emergence of a $U(1)$ symmetry has also been observed in other Rydberg models~\cite{Zhou_2025}, but these are not straightforward to implement experimentally.

For circumferences $l_y=1$ and $l_y=2$, we focus on the two longest open cylinders considered for the FSS analysis presented above, with $l_x=7$ and $l_x=10$, respectively. For $l_y=3$, we consider the longest accessible cylinder with $l_x=5$ ($N_x=15$ atoms).  
We choose values of $\Delta/U$ close to the transition point $\Delta_{{\rm c}, (l_x,l_y)}/U$, which we estimate by the point where the order parameter $\langle m^3+m^{3\dagger}\rangle $ vanishes. We find $\Delta_{{\rm c}, (l_x,l_y)}\approx3.15876$, $3.158097,$ and $3.15879$ for the three respective geometries. The chosen values of $\Delta/U$ correspond to the following distances from criticality: $\bar{\Delta} = \Delta/U-\Delta_{{\rm c}, (l_x,l_y)}/U \approx 0.0027$, $-0.0002$, and $-0.0004$, for $l_y=1,2,$ and $3$, respectively.

In Fig.~\ref{fig:Sampling_open_cylinders_App}, we present the distribution of $m=\rho e^{i\phi}$ in the complex plane, as well as the marginal distributions for its amplitude and phase, $P(\rho)$ and $P(\phi)$. These marginal distributions are compared with the analytical predictions of Eq.~\eqref{eq:marginal_probab_phi} and Eq.~\eqref{eq:marginal_probab_rho} (red dashed line), respectively, evaluated using histogram binning with a chosen bin width. In all cases, the analytical fit agrees well with a Gaussian kernel density estimate of the data, which provides a smoother and more reliable characterization of the underlying distribution than binning alone.

These results confirm that the radial distribution compatible with a simple local effective potential $V(\rho) =- \mu \rho^2 + \lambda \rho^4$, with $\mu, \lambda > 0$, while the angular distribution is well described by a potential of the form $V(\phi) = g_3 \rho_0^3 \cos(3\phi) + g_6 \rho_0^6 \cos(6\phi)$. Here, the term $g_3$ arises due to deviations from the critical point (i.e., $g_3 \propto \bar{\Delta} \neq 0$), as confirmed by the fitted values of $g_3$, whose sign matches that of $\bar{\Delta}$.

\begin{figure}
    \centering
    \includegraphics[width=\columnwidth]{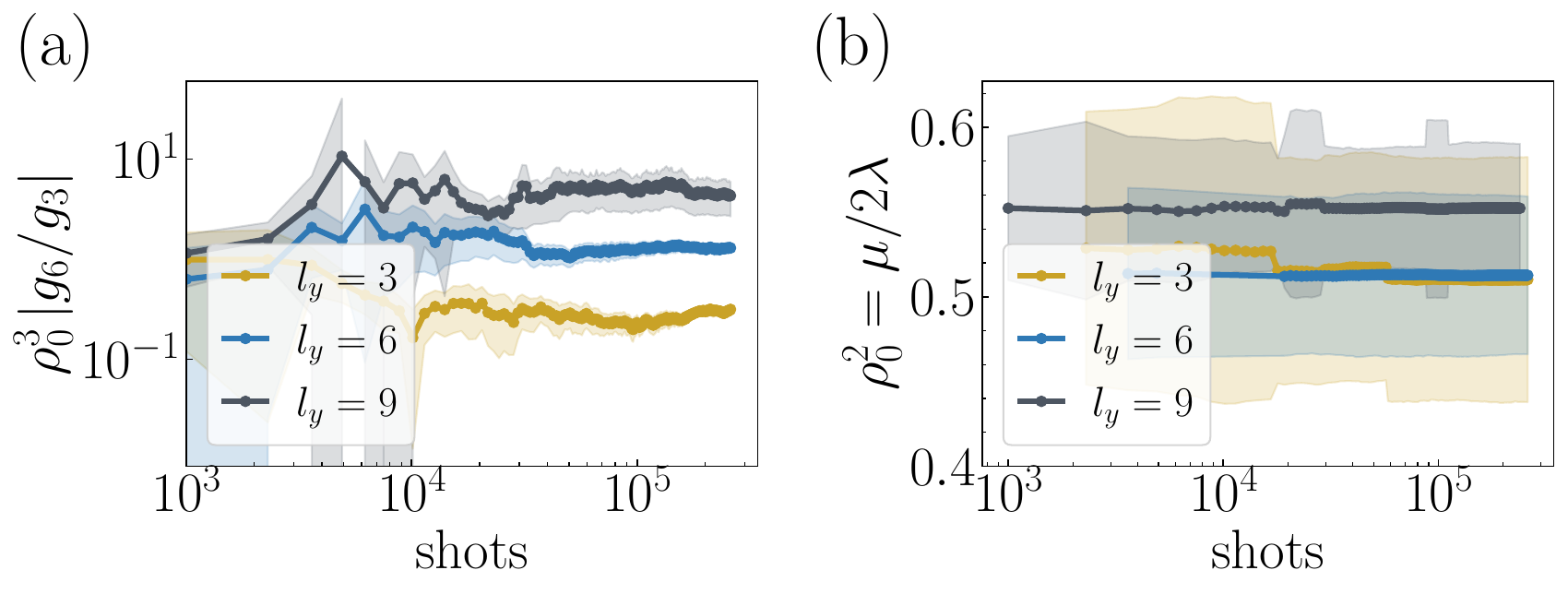}
    \caption{Estimated values of (a) $\rho_0^3 |g_6 / g_3|$ and (b) $\rho_0^2$ as a function of the number of measurement shots. Solid lines indicate the fitted values, and shaded regions represent the corresponding $\pm \sigma$ confidence intervals. The estimates are obtained from fits of the marginal distributions $P(\phi)$ and $P(\rho)$ in Fig.~\ref{fig:Sampling_open_cylinders_App}, for the three cylinders with $l_y=1,2,$ and $3$, using Eqs.~\eqref{eq:marginal_probab_phi} and~\eqref{eq:marginal_probab_rho}, respectively. }
    \label{fig:Sampling_open_cylinders_App_fit_shots_g3g6}
\end{figure}

In Fig.~\ref{fig:Sampling_open_cylinders_App_fit_shots_g3g6}(a), we show the ratio $\rho_0^3 |g_6 / g_3|$, estimated from the radial distribution $P(\rho)$, as a function of the number of shots for each $l_y$. The estimates converge after approximately $10^5$ shots, indicating that this number is sufficient to reliably extract the parameter. We observe that $\rho_0^3 |g_6 / g_3|$ increases for the different cylinders of increasing $l_y$. 
Analytically, this scaling is expected to arise from the dependence of $|g_6 / g_3|$ on $l_y$, while $\rho_0^2$ is expected to be independent of it. This is supported by Fig.~\ref{fig:Sampling_open_cylinders_App_fit_shots_g3g6}(b), which shows that $\rho_0^2$, estimated from the radial distribution $P(\rho)$, exhibits only a weak dependence on $l_y$.
The observed increase of $|g_6 / g_3|$ across the three cylinders may originate from two combined effects: the decrease in the aspect ratio $l_x/l_y$ ($=7,5, 1.7$, respectively) which enhances finite-size effects, and the increase in $l_y$ for which the $\cos(6\phi)$ anisotropy term is expected to become more relevant.

\subsection{Including the interaction up to the fifth-nearest neighbors}
\begin{figure}[b]
    \centering
    \includegraphics[width=\columnwidth]{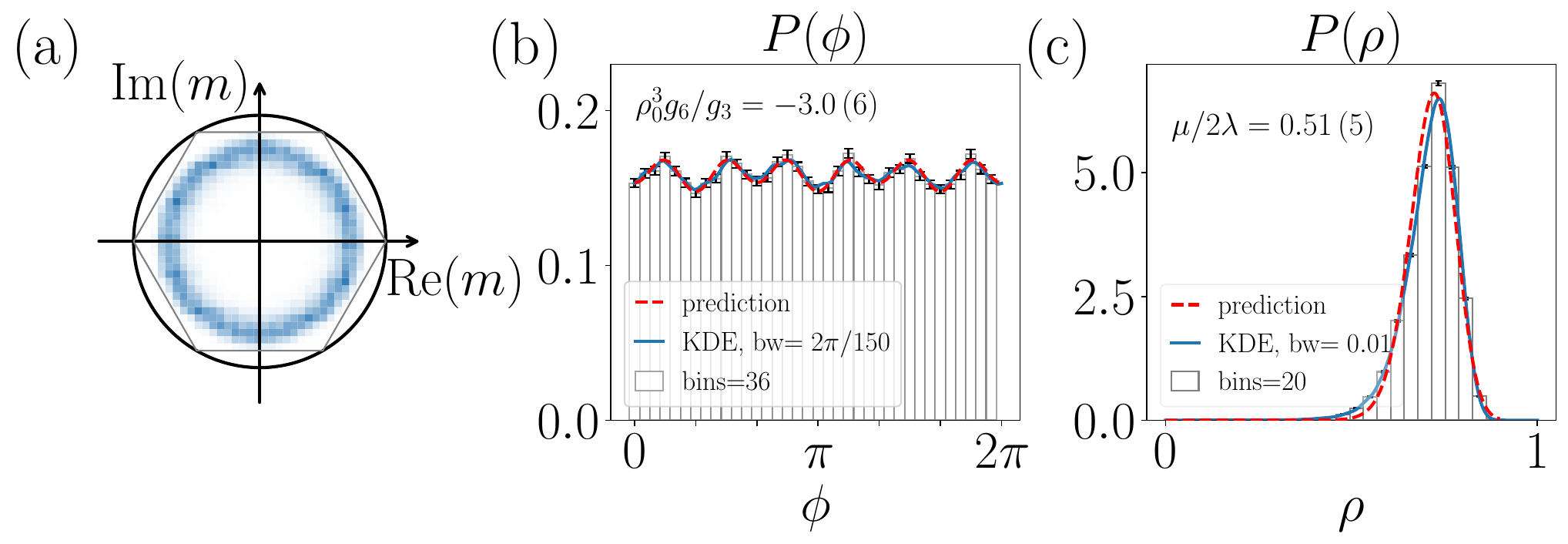}
    \caption{Sampling of the staggered magnetization $m$ [Eq.~$(2)$] for a cylinder with $l_y=2$ and $l_x=10$ at $\Delta/U=3.1796$, including interactions up to the fifth-nearest neighbors (cf. second column in Fig.~\ref{fig:Sampling_open_cylinders_App} for interactions up to the third-nearest neighbors).  (a) Distribution of $m$ over $10^5$ snapshots in the occupation basis, restricted to the bulk.  (b) Corresponding angular distribution,  fitted according to Eq.~\eqref{eq:marginal_probab_phi} (red dashed line).  (c) Corresponding radial distribution, fitted according to Eq.~\eqref{eq:marginal_probab_rho} (red dashed line). Blue lines in (b)-(c) indicate a Gaussian kernel density estimate. Results obtained using a finite MPS with bond dimension $\chi=250$ for $l_y=2$.}
    \label{fig:5NN}
\end{figure}
Here, we consider the open cylinder with $l_y = 2$ and $l_x = 5$, and we show that extending the range of interactions beyond the third-nearest neighbor---up to the fifth-nearest neighbor---does not affect the emergence of U($1$) symmetry at criticality. In the classical limit ($\Omega=0$), the extended interaction tails induce a shift in the critical point, from $\Delta/U\approx3.158$ to $\Delta/U\approx3.1796$. Nevertheless, as demonstrated in Fig.~\ref{fig:5NN}, the two-dimensional and marginal distributions of the staggered magnetization near criticality remain consistent with those obtained in the shorter-range interaction case discussed previously [cf. second column in Fig.~\ref{fig:Sampling_open_cylinders_App}].

\section{Open systems including the interaction up to the fifth-nearest neighbors}\label{sec:App_open_systems}

Here, we investigate the experimentally feasible setup shown in Fig.~3(c)–(f), incorporating interactions up to the fifth-nearest neighbor. We demonstrate that the staggered magnetization $m$, computed from $10^4$ snapshots in the occupation basis, serves as an experimentally accessible observable to distinguish between the two ordered phases with $\mathbb{Z}_3$ symmetry and the critical regime featuring an emergent $U(1)$ symmetry.

\begin{figure}
    \centering
    \includegraphics[width=\columnwidth]{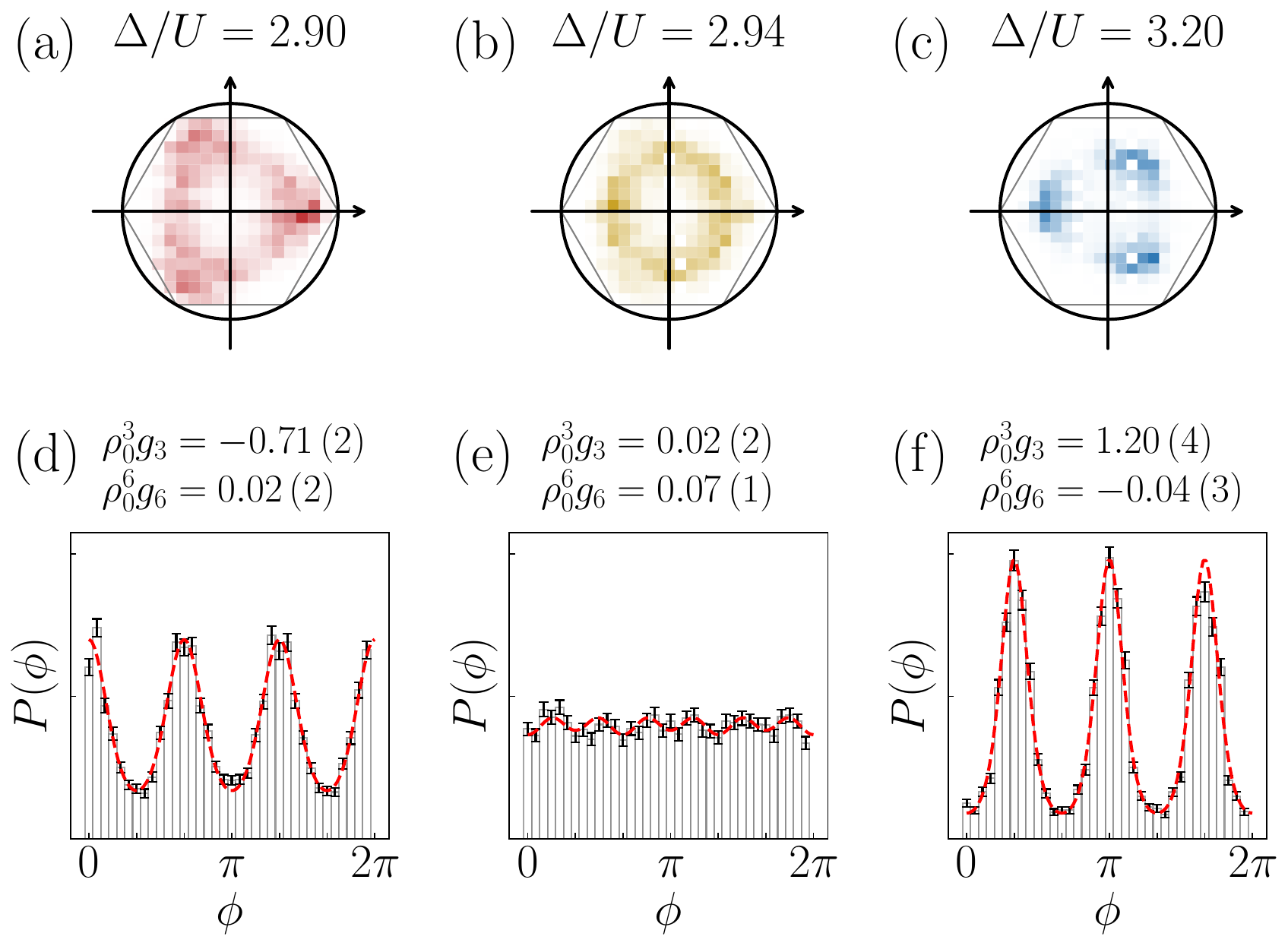}
    \caption{Statistics of the staggered magnetization for the experimentally feasible setup shown in Fig.~3(c)–(f), incorporating interactions up to the fifth-nearest neighbor. (First column) $\Delta/U = 2.9$, in the $1/3$ phase; (Second column) $\Delta/U = 2.94$, near the transition; and (Third column) $\Delta/U = 3.2$, in the $2/3$ phase. For each cases, we show: (First row) the distribution of $m$ obtained from $10^5$ snapshots in the occupation basis, restricted to the bulk; and (Second row) the corresponding angular distribution, fitted according to Eq.~\eqref{eq:marginal_probab_phi} (red dashed line). Results obtained using a finite MPS with bond dimension $\chi=300$.}
    \label{fig:Open_sys_SM}
\end{figure}

In Fig.~\ref{fig:Open_sys_SM}, we present the angular probability distribution $P(\phi)$ of $m$ as the system undergoes the phase transition between the $1/3$ and $2/3$ phases.
In the ordered phases, $P(\phi)$ exhibits three distinct maxima: at $\phi = 0, 2\pi/3,$ and $ 4\pi/3$ for $\Delta/U = 2.9$ ($1/3$ phase), and at $\phi = \pi/3, \pi,$ and $ 5\pi/3$ for $\Delta/U = 3.2$ ($2/3$ phase). At the transition ($\Delta/U \approx 2.92$), an approximately U($1$)-symmetric angular distribution emerges, which is thus robust to extension of the interaction tails [cf. Fig.~3(d)–(f)].
In particular, from the fit of these angular distributions using Eq.~\eqref{eq:marginal_probab_phi}, we can estimate the coefficients of the effective field theory presented in App.~\ref{sec:SM_appA}. 
We consistently find that $|g_3/g_6|\gg 1$ in the $\mathbb{Z}_3$-ordered phases and $g_3$ changes sign across the transition---being negative in the $1/3$ phase and positive in the $2/3$ phase. 

\section{Phase transition for varying $\Omega/U$}\label{sec:SM_vary_Omega}
\begin{figure}
    \centering
    \includegraphics[width=\columnwidth]{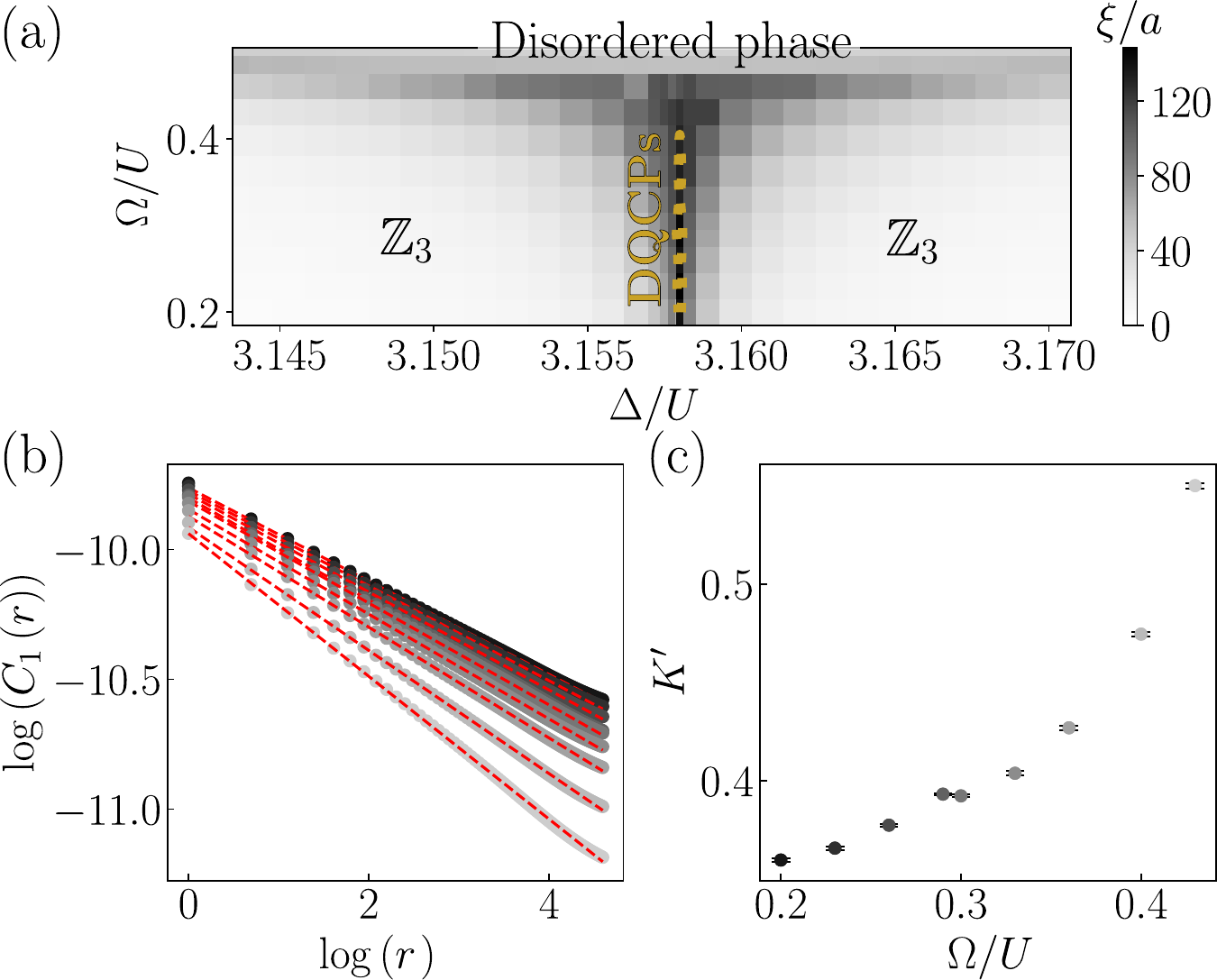}
    \caption{Analysis of the transition between the $1/3$ and the $2/3$ phases as a function of $\Omega/U$, for an infinitely long cylinder with $l_y=1$. (a) Correlation length $\xi/a$ as a function of $\Delta/U$ and $\Omega/U$. 
    (b) Power-law decay of the two-point correlation function $\langle m(r) m(0)\rangle \approx C_1(r)=\rho^2_0 \langle e^{i\phi(r)}  e^{-i\phi(0)} \rangle$ at criticality for $\Omega/U\leq 0.43$. The colors refer to the $\Omega/U$  values shown in (c). (c) Luttinger parameter $K'$, extracted from linear fits shown (red lines) in (b), as a function of $\Omega/U$. Results obtained using a finite MPS with bond dimension $\chi=150$.}
    \label{fig:iMPS_33_vary_OmegaU_paramters_th}
\end{figure}
Here, we consider the infinitely long cylinder with the smallest circumference ($l_y=1$) and analyze the phase diagram near the transition between the $1/3$ and $2/3$ phases as a function of $\Omega/U$. We find that the two ordered phases are separated by a line of DQCPs for small values of $\Omega/U$ ($\Omega/U\lesssim 0.43$) [see dashed yellow line in Fig.~\ref{fig:iMPS_33_vary_OmegaU_paramters_th}(a)], where the two-point correlation function $C_1(r)$ exhibits power-law decay [Fig.~\ref{fig:iMPS_33_vary_OmegaU_paramters_th}(b)]. The location of the transition, i.e., the position of the maximum in the correlation length $\xi$, remains unchanged with varying $\Omega/U$. However, $\Omega/U$ affects the Luttinger parameter $K'$. Specifically, the exponent governing the power-law decay of the two-point correlation function $\langle m(r) m(0) \rangle$ at criticality, which is given by $K'/2$, increases with increasing $\Omega/U$ [Fig.~\ref{fig:iMPS_33_vary_OmegaU_paramters_th}(c)].

\section{Estimates of feasible experimental parameter ranges}\label{App:exp_paramters}

Here, we provide realistic experimental parameters for a possible state preparation of the configuration shown in Fig.~3(c), consisting of 75 atoms. 
We consider a Rydberg atom array with a van der Waals interaction coefficient $C_6 = 2\pi \cdot 650$ GHz~\textmu m$^6$ and a lattice spacing of $a=7.5$ ~\textmu m, yielding a nearest-neighbor interaction strength of $U \approx 2\pi \cdot 3.7$ MHz. The resulting total system size is approximately $58.5 \times 48.75$  ~\textmu m$^2$. At the end of the adiabatic state preparation, the detuning and Rabi frequency should be set to $\Delta = 2.94 U \approx 2\pi \cdot 11$~MHz and $\Omega = 0.33U \approx 2\pi \cdot 1.2$~MHz, respectively, while a larger Rabi frequency up to $\Omega = U \approx 2\pi \cdot 3.7$~MHz can be applied during the ramp to access the disordered phase.

These parameters are well within the range of current experimental capabilities. Typical Rydberg setups achieve $C_6 \approx 600$–$650$ GHz~\textmu m$^6$ with lattice spacings above $4$~\textmu m. The available detuning bandwidth is around $2\pi \cdot 40$~MHz, while Rabi frequencies up to $2\pi \cdot 6$~MHz are routinely attainable for arrays up to around $100$~\textmu m in width. Furthermore, current experiments with $300$–$500$ atoms can reach coherence times of a few microseconds ($T_2^* \approx 2$–$4$~\textmu s), allowing dynamical evolutions of up to $\sim 3$~\textmu s.

Since the gap of the DQCP closes polynomially with the system size, the evolution time required to prepare this critical state also increases accordingly. Consequently, the finite coherence time of the experiment will set an upper bound on the accessible system size. A detailed numerical analysis will be necessary to quantify this limitation, which we plan to carry out in future work.

\end{document}